\documentclass[aps,prd,preprintnumbers,groupedaddress,nofootinbib,amssymb,eqsecnum,notitlepage]{revtex4-1}
\usepackage{graphicx}
\usepackage{bm}
\usepackage{amsmath}
\usepackage{color}
\usepackage{amsfonts}
\usepackage{here}
\usepackage{graphicx}
\usepackage{amsmath,amsthm,amssymb}
\usepackage{bm}
\allowdisplaybreaks[1]

\usepackage{amsfonts}
\usepackage{dcolumn}
\usepackage{hyperref}

\begin{document}
\newcommand{\newc}{\newcommand}
\newc{\tif}{\tilde{f}}
\newc{\tih}{\tilde{h}}
\newc{\tip}{\tilde{\phi}}
\newc{\tiA}{\tilde{A}}

\newcommand{\ben}{\begin{eqnarray}}
\newcommand{\een}{\end{eqnarray}}
\newc{\be}{\begin{equation}}
\newc{\ee}{\end{equation}}
\newc{\ba}{\begin{eqnarray}}
\newc{\ea}{\end{eqnarray}}
\newc{\bea}{\begin{eqnarray*}}
\newc{\eea}{\end{eqnarray*}}
\newc{\D}{\partial}
\newc{\ie}{{\it i.e.} }
\newc{\eg}{{\it e.g.} }
\newc{\etc}{{\it etc.} }
\newc{\etal}{{\it et al.}}
\newcommand{\nn}{\nonumber}
\newc{\ra}{\rightarrow}
\newc{\lra}{\leftrightarrow}
\newc{\lsim}{\buildrel{<}\over{\sim}}
\newc{\gsim}{\buildrel{>}\over{\sim}}
\newc{\aP}{\alpha_{\rm P}}
\newc{\dphi}{\delta\phi}
\newc{\da}{\delta A}
\newc{\tp}{\dot{\phi}}
\newc{\Mpl}{M_{\rm pl}}
\newc{\bem}{\beta_m}
\newc{\beG}{\beta_G}
\newc{\beA}{\beta_A}

\title{Cosmology in scalar-vector-tensor theories}

\author{
Lavinia Heisenberg$^{1}$, 
Ryotaro Kase$^{2}$, and 
Shinji Tsujikawa$^{2}$}

\affiliation{
$^1$Institute for Theoretical Studies, ETH Zurich, 
Clausiusstrasse 47, 8092 Zurich, Switzerland\\
$^2$Department of Physics, Faculty of Science, 
Tokyo University of Science, 1-3, Kagurazaka,
Shinjuku-ku, Tokyo 162-8601, Japan}

\date{\today}

\begin{abstract}

We study the cosmology on the Friedmann-Lema\^{i}tre-Robertson-Walker (FLRW) 
background in scalar-vector-tensor theories with a broken $U(1)$ gauge symmetry. 
For parity-invariant interactions arising in 
scalar-vector-tensor  theories with second-order equations 
of motion, we derive conditions for the absence of ghosts and 
Laplacian instabilities associated with tensor, vector, 
and scalar perturbations at linear order. 
This general result is applied to the computation 
of the primordial tensor power spectrum 
generated during inflation as well as to 
the speed of gravity relevant to the late-time cosmic acceleration.
We also construct a concrete inflationary model in which 
a temporal vector component $A_0$ contributes to the 
dynamics of cosmic acceleration besides a scalar field $\phi$ 
through their kinetic mixings.
In this model, we show that all the stability conditions 
of perturbations can be consistently satisfied during 
inflation and subsequent reheating.

\end{abstract}


\maketitle

\section{Introduction}
\label{introsec}

Despite the tremendous progress of observational cosmology 
over the past two decades, there are several unsolved issues 
in theoretical cosmology. 
The observations of Cosmic Microwave Background (CMB) \cite{CMB} and 
supernovae type Ia \cite{SN} have shown that our Universe exhibited 
two stages of cosmic acceleration: inflation and dark energy. 
Moreover, we know that dark matter played a crucial role for the 
large-scale structure formation \cite{LSS}. 
The existing problems of inflation, dark energy, and dark matter 
imply that there may be some extra degrees of 
freedom (DOFs) beyond the paradigms of standard model 
of particle physics and General Relativity (GR) \cite{review}.

A scalar field $\phi$ can be a natural candidate for addressing 
such problems. In theories aiming to unify quantum field theory and GR, 
the scalar field can generally have direct couplings to 
gravity. A dilaton field arising in string theory is one 
of such examples, in which case there is a nonminimal coupling of 
the form $F(\phi)R$ with the Ricci scalar $R$ \cite{Gas}.
One can also consider a derivative interaction in which the 
field kinetic energy $-\partial_{\mu}\phi \partial^{\mu}\phi/2$ 
is directly coupled to $R$ \cite{Amendola}. 
In such cases, however, the theories generally contain derivatives 
higher than second order, so they are plagued 
by the problem of so-called Ostrogradski instabilities \cite{Ostro}.
It is possible to keep the equations of motion up to second order 
by adding counter terms in the Lagrangian to eliminate higher-order 
derivatives \cite{Deffayet}. The most general scalar-tensor theories with 
second-order equations of motion are dubbed Horndeski 
theories \cite{Horndeski,Horn2,Horn3}, 
which have been widely applied to the construction of 
viable models of inflation and dark 
energy \cite{Ginf,Burrage,KYY,DT11,GSami,DT10,Hornap,CCPS,Bellini,Heisenberg:2014kea}.

A vector field $A_{\mu}$ can also be the source for 
cosmic acceleration. If the vector field coupled to gravity 
respects the $U(1)$ gauge symmetry as well as the Lorentz invariance, 
it is not possible to construct nontrivial derivative interactions 
such as those appearing in scalar Horndeski theories \cite{nogo}. 
The vector field with a broken $U(1)$ symmetry
(including a massive Proca field) allows 
Galileon-type derivative and nonminimal couplings to gravity.
Unlike scalar-tensor theories, there are also new interactions 
arising from intrinsic vector modes \cite{Heisenberg}. 
Most general vector-tensor theories with second-order 
equations of motion are dubbed generalized Proca (GP) 
theories \cite{Heisenberg,Allys,Jimenez}. 
The applications of GP theories to 
dark energy \cite{Tasinato,GPcosmo1,GPcosmo2,GPcosmo3} and spherically 
symmetric objects \cite{GPsc,GPBH1,GPBH2,GPBH3,GPBH4,GPBH5} 
were extensively performed in the literature.

It is possible to unify Horndeski and GP 
theories in the form of scalar-vector-tensor (SVT) theories. 
In Ref.~\cite{Heisenberg18}, the action of SVT theories with second-order 
equations of motion was constructed by keeping the $U(1)$ gauge symmetry 
or by abandoning it. In the gauge-invariant setup
the longitudinal component of a vector field $A_{\mu}$ does not propagate, 
so a scalar field $\phi$ is the only scalar propagating DOF besides two 
transverse vector modes and two tensor polarizations \cite{Heisenberg18}.
In this case, two of present authors found a new type of hairy black hole 
solutions in a static and spherically symmetric 
background \cite{HT18} (see also Refs.~\cite{Cha,Pedro}), which are 
stable against odd-parity perturbations under certain bounds of coupling 
constants \cite{HKT18}.

If we try to apply SVT theories to cosmology, the $U(1)$ invariant theories 
do not allow the existence of a time-dependent vector field relevant to 
the dynamics on the FLRW background.
In this case, the vector field needs to be promoted to a non-abelian 
gauge field with a broken $SU(2)$ symmetry \cite{noAbe1,noAbe2}, 
which we will not consider in this paper.
In SVT theories with broken $U(1)$ gauge invariance, the 
time-dependent temporal vector component $A_0(t)$ can play 
a role for the background cosmology besides a scalar field $\phi (t)$ \cite{Heisenberg18}.
It is of interest to apply such new theories
to the dynamics of inflation and dark energy. 
In particular, there are six propagating DOFs (two scalars, two vectors, 
and two tensors) in SVT theories with broken $U(1)$ symmetry, so 
we need to study whether any of the propagating DOFs are 
plagued by instability problems.

In this paper, we derive conditions for the absence of 
ghosts and Laplacian instabilities of linear cosmological perturbations 
in the presence of most general $U(1)$ broken SVT interactions with  second-order equations of motion and with parity invariance. 
In Sec.~\ref{modelsec}, we revisit the action of $U(1)$ broken SVT theories and obtain 
the background equations of motion on the flat FLRW spacetime. 
In Sec.~\ref{tensorsec}, we compute the second-order action of tensor perturbations 
and apply it to the speed of gravitational waves relevant to the late-time 
cosmic acceleration and to 
the calculation of the primordial tensor power spectrum generated during inflation.
In Secs.~\ref{vectorsec} and \ref{scalarsec}, we obtain conditions for avoiding 
ghosts and Laplacian instabilities of vector and scalar perturbations 
by deriving their second-order actions. 
In Sec.~\ref{apsec}, we construct a concrete inflationary model in the framework 
of SVT theories and show that all the stability conditions can be consistently 
satisfied during inflation and reheating.
Sec.~\ref{concludesec} is devoted to conclusions. 
Throughout the paper, we use the natural unit in which the speed of 
light $c$ is equivalent to 1.

\section{$U(1)$ broken SVT theories and 
background equations of motion}
\label{modelsec}

In Ref.~\cite{Heisenberg18}, the action of SVT theories with broken $U(1)$ symmetry was constructed by unifying Horndeski theories with GP theories. 
In this paper, we focus on new interactions arising in SVT theories with second-order equations of motion and apply 
them to the cosmological dynamics
on the flat FLRW background. 
The $U(1)$ broken SVT theories consist of a vector field 
$A_{\mu}$ and a scalar field $\phi$, both of which have 
direct couplings to gravity.

\subsection{SVT theories with broken $U(1)$ symmetry}
 
We define a field strength $F_{\mu\nu}$ of the vector field $A_{\mu}$, 
its dual $\tilde{F}^{\mu\nu}$, and 
a symmetric tensor $S_{\mu\nu}$, as
\be
F_{\mu\nu}=\nabla_\mu A_\nu-\nabla_\nu A_\mu\,,
\qquad 
\tilde{F}^{\mu\nu}=\frac{1}{2}
\mathcal{E}^{\mu\nu\alpha\beta}F_{\alpha\beta}\,,
\qquad
S_{\mu\nu}=\nabla_\mu A_\nu+\nabla_\nu A_\mu\,,
\ee
where $\nabla_{\mu}$ represents a covariant derivative operator, 
and $\mathcal{E}^{\mu\nu\alpha\beta}$ is the anti-symmetric Levi-Civita tensor 
satisfying the normalization $\mathcal{E}^{\mu\nu\alpha\beta}
\mathcal{E}_{\mu\nu\alpha\beta}=-4!$. 
While neither $F_{\mu \nu}$ nor $\tilde{F}^{\mu\nu}$ affects the cosmological
background dynamics with a purely temporal component, this is not the case for $S_{\mu \nu}$.
We introduce another tensor for convenience in form of an effective metric 
constructed from possible combinations of 
$g_{\mu \nu}$, $A_{\mu}$, and $\nabla_{\mu}\phi$, i.e., 
\be
\mathcal{G}_{\mu\nu}^{h_{n}} 
=h_{n1}(\phi,X_i)g_{\mu\nu}+h_{n2}(\phi,X_i)
\nabla_\mu \phi \nabla_\nu \phi
+h_{n3}(\phi,X_i)A_\mu A_\nu+h_{n4}(\phi,X_i)A_\mu 
\nabla_\nu \phi\,,
\label{effmet}
\ee
where $g_{\mu \nu}$ is the four-dimensional 
spacetime metric, and 
\be
X_1=-\frac{1}{2} \nabla_{\mu} \phi \nabla^{\mu} \phi\,,
\qquad
X_2=-\frac{1}{2} A^{\mu} \nabla_{\mu} \phi \,,
\qquad 
X_3=-\frac{1}{2} A_{\mu} A^{\mu}\,,
\ee
with $i=1,2,3$. 
As we will see below, $\mathcal{G}_{\mu\nu}^{h_{n}}$ 
appears in the fifth-order Lagrangian 
of SVT theories\footnote{As was pointed out 
in Ref.~\cite{Heisenberg18}, the explicit dependence on all the $h_{nj}$ functions needs an additional caution, since an arbitrary dependence on a general background will introduce dynamics for the temporal component of the vector field. In order for this not to happen, the dependence of $\mathcal{M}_5^{\mu\nu}$ would need to be restricted to $X_1$ and similarly the dependence of $\mathcal{N}_5^{\mu\nu}$ to $X_3$ and so on. We leave them here as general functions since the background symmetries are not oblivious to this fact, but for a more general background this would need to be taken into account.\label{foot1}}, 
so that the subscript $n$ represents $n=5$.
The new action arising in SVT theories with broken 
$U(1)$ symmetry is given by \cite{Heisenberg18}
\be
\mathcal{S}_{\rm SVT}=
\int d^4x \sqrt{-g}\,\sum_{n=2}^6\mathcal{L}_n\,,
\label{action}
\ee
with the Lagrangians
\ba
\label{genLagrangianSVTnoGauge}
\mathcal{L}_{2} &=&
f_2(\phi,X_1,X_2,X_3,F,Y_1,Y_2,Y_3)\,, \nonumber\\
\mathcal{L}_{3} &=&
f_{3}(\phi,X_3)g^{\mu\nu}S_{\mu\nu}
+\tilde{f}_{3}(\phi,X_3)A^{\mu}A^{\nu} S_{\mu\nu}\,,\nonumber\\
\mathcal L_{4} & = & 
f_{4}(\phi,X_3)R+f_{4,X_3}(\phi,X_3) \left[ 
(\nabla_\mu A^\mu)^2-\nabla_\mu A_\nu \nabla^\nu A^\mu \right]\,, \nonumber\\
\mathcal L_{5} & = & 
f_5(\phi,X_3)G^{\mu\nu} \nabla_{\mu}A_{\nu}
-\frac{1}{6}f_{5,X_3}(\phi,X_3) 
\left[ (\nabla_{\mu} A^{\mu})^3
-3\nabla_{\mu} A^{\mu}
\nabla_{\rho}A_{\sigma} \nabla^{\sigma}A^{\rho}
+2\nabla_{\rho}A_{\sigma} \nabla^{\gamma}
A^{\rho} \nabla^{\sigma}A_{\gamma}\right] \nonumber\\
&+&\mathcal{M}_5^{\mu\nu}\nabla_\mu \nabla_\nu\phi
+\mathcal{N}_5^{\mu\nu}S_{\mu\nu}\,, 
\nonumber\\
\mathcal{L}_{6} & = &
f_6(\phi,X_1)L^{\mu\nu\alpha\beta}F_{\mu\nu}F_{\alpha\beta}
+\mathcal{M}_6^{\mu\nu\alpha\beta}\nabla_\mu\nabla_\alpha \phi\nabla_\nu\nabla_\beta\phi+\tilde{f}_6(\phi,X_3)L^{\mu\nu\alpha\beta}F_{\mu\nu}F_{\alpha\beta}+ \mathcal{N}_6^{\mu\nu\alpha\beta}S_{\mu\alpha}S_{\nu\beta}\,,
\ea
where $R$ and $G^{\mu \nu}$ are the Ricci scalar and the 
Einstein tensor, respectively, and 
$f_{4,X_3}=\partial f_4/\partial X_3$, 
$f_{5,X_3}=\partial f_5/\partial X_3$.
The double dual Riemann tensor $L^{\mu \nu \alpha \beta}$ 
is defined by 
\be
L^{\mu\nu\alpha\beta}=\frac{1}{4}
\mathcal{E}^{\mu\nu\rho\sigma}
\mathcal{E}^{\alpha\beta\gamma\delta} R_{\rho\sigma\gamma\delta}\,,
\ee
where $R_{\rho\sigma\gamma\delta}$ is the Riemann tensor. 
We also used the following notations:
\be
F=-\frac{1}{4}F_{\mu\nu}F^{\mu\nu}\,,\qquad
Y_1=\nabla_\mu \phi \nabla_\nu \phi\,F^{\mu\alpha}F^\nu{}_\alpha\,,
\qquad 
Y_2=\nabla_\mu\phi\, A_\nu F^{\mu\alpha}F^\nu{}_\alpha\,, 
\qquad
Y_3=A_\mu A_\nu F^{\mu\alpha}F^\nu{}_\alpha\,,
\ee
which correspond to the interactions arising from 
pure vector modes.

The 2-rank tensors $\mathcal{M}^{\mu\nu}_5$ and 
$\mathcal{N}^{\mu\nu}_5$ in ${\cal L}_5$, which 
encode intrinsic vector interactions, are given by 
\be
\mathcal{M}^{\mu\nu}_5
=\mathcal{G}_{\rho\sigma}^{h_{5}} 
\tilde{F}^{\mu\rho}\tilde{F}^{\nu\sigma}\,,\qquad 
\mathcal{N}^{\mu\nu}_5
=\mathcal{G}_{\rho\sigma}^{\tilde{h}_{5}} 
\tilde{F}^{\mu\rho}\tilde{F}^{\nu\sigma}\,,
\ee
where the functions $h_{5j}$ and $\tilde{h}_{5j}$ ($j=1,2,3,4$) 
appearing in ${\cal G}_{\rho \sigma}^{h_5}$ and 
${\cal G}_{\rho \sigma}^{\tilde{h}_5}$ 
are functions of $\phi$ 
and $X_1, X_2, X_3$. 
The Lagrangian ${\cal L}_6$ also corresponds to 
the interactions of intrinsic vector modes.
The 4-rank tensors $\mathcal{M}_6^{\mu\nu\alpha\beta}$ 
and $\mathcal{N}^{\mu\nu\alpha\beta}_6$ are defined by 
\be
\mathcal{M}^{\mu\nu\alpha\beta}_6
=2f_{6,X_1} (\phi, X_1) 
\tilde{F}^{\mu\nu}\tilde{F}^{\alpha\beta}\,,
\qquad
\mathcal{N}^{\mu\nu\alpha\beta}_6
=\frac12\tilde{f}_{6,X_3} (\phi, X_3) 
\tilde{F}^{\mu\nu}\tilde{F}^{\alpha\beta}\,.
\ee
The functions $f_3, \tilde{f}_3, f_4, f_5, \tilde{f}_6$ depend 
on $\phi$ and $X_3$, whereas $f_6$ has the dependence of 
$\phi$ and $X_1$. The function $f_2$ contains the dependence 
of $\phi, X_1, X_2, X_3, F, Y_1, Y_2, Y_3$. In $f_2$, we 
do not take into account the parity-violating 
term $\tilde{F}=-F_{\mu \nu} \tilde{F}^{\mu \nu}/4$ from \cite{Heisenberg18}. 

The action of GP theories 
(which is given by Eqs.~(2.2)-(2.6) of Ref.~\cite{GPcosmo2}) 
can be recovered by using the correspondence
$\phi \to 0, X_{1,2} \to 0, X_3 \to X, Y_{1,2} \to 0, Y_3 \to Y$, 
$f_2 \to G_2(X,F,Y)$,
$2f_3 \to G_3(X)$, $\tilde{f}_3 \to 0$, 
$f_4 \to G_4(X)$, $f_5 \to G_5(X)$, 
$h_{5j} \to 0$, $\tilde{h}_{51} \to -g_5(X)/2$, 
$\tilde{h}_{52}, \tilde{h}_{53}, \tilde{h}_{54}\to 0$, 
$f_6 \to 0$, and 
$4\tilde{f}_6 \to G_6(X)$ in the action (\ref{action}).

We note that the full action of SVT theories with second-order 
equations of motion is given by ${\cal S}=
{\cal S}_{\rm SVT}+{\cal S}_{\rm ST}$, where 
${\cal S}_{\rm ST}$ is the action of scalar-tensor 
Horndeski theories with the Lagrangians (2.1)-(2.4) 
of Ref.~\cite{Horn3}.
Since we are interested in the effect of new interactions 
${\cal S}_{\rm SVT}$ on the cosmological dynamics,  
we focus on $U(1)$ broken SVT theories 
given by the action (\ref{action}).
In such theories, there are six propagating 
DOFs in total (two tensors, two vectors, and two scalars) 
on the flat FLRW background.
In Secs.~\ref{tensorsec}-\ref{scalarsec},
we study the propagation of tensor, vector, 
and scalar perturbations in turn. 
In Sec.~\ref{apsec}, we apply our $U(1)$ broken SVT 
theories to the inflationary cosmology.

\subsection{Background equations of motion}

To derive the equations of motion on the flat FLRW background, we begin with the line element
\be
ds^2=-N^2(t) dt^2+a^2(t) \delta_{ij}dx^i dx^j\,,
\label{FLRW}
\ee
where $N(t)$ is the lapse and $a(t)$ is the scale factor.
We also consider the configuration of a time-dependent 
scalar field $\phi(t)$ and a vector field $A_{\mu} (t)$ 
given by 
\be
A_{\mu}(t)=\left( A_0(t) N(t), 0, 0, 0 \right)\,,
\ee
where $A_0(t)$ is a time-dependent temporal vector component.
The quantities $F, Y_1, Y_2, Y_3$ vanish on the spacetime 
metric (\ref{FLRW}), so they do not contribute to 
the background equations of motion.
Moreover, the Lagrangian ${\cal L}_6$ and the interactions 
proportional to $\mathcal{M}^{\mu\nu}_5$ and $\mathcal{N}^{\mu\nu}_5$ 
in ${\cal L}_5$ do not affect the background cosmology either. 
The quantities $X_1$, $X_2$, $X_3$ are given, respectively, by 
\be
X_1=\frac{\dot{\phi}^2}{2N^2}\,,\qquad 
X_2=\frac{\dot{\phi}A_0}{2N}\,,\qquad 
X_3=\frac{A_0^2}{2}\,,
\ee
where a dot represents a derivative with respect to $t$.
We compute the action (\ref{action}) on the spacetime metric
(\ref{FLRW}) and vary it with respect to $N$, $a$, $\phi$, 
and $A_0$. Setting $N=1$ at the end, we obtain the following 
equations of motion on the flat FLRW background:
\ba
\hspace{-0.3cm}
& &
6f_4H^2+f_2-\dot{\phi}^2 f_{2,X_1}-\frac{1}{2} \dot{\phi} 
A_0 f_{2,X_2}+6H \left( \dot{\phi}f_{4,\phi}-HA_0^2 f_{4,X_3} \right)
+2A_0 H^2 \left( 3\dot{\phi}f_{5,\phi}-A_0^2 H 
f_{5,X_3} \right)=0\,,\label{back1}\\
\hspace{-0.3cm}
& &
2f_4 \left( 2\dot{H}+3H^2 \right)+f_2+2\dot{A}_0 A_0^2 
\left( f_{3,X_3}+\tif_3 \right)+2\dot{\phi}A_0 f_{3,\phi}
+2\left( \ddot{\phi}+2H \dot{\phi} \right)f_{4,\phi}
-2A_0 \left[ A_0 (2\dot{H}+3H^2)
+2\dot{A}_0 H \right]f_{4,X_3}
\nonumber \\
\hspace{-0.3cm}
& &
+2\dot{\phi}\dot{A}_0 A_0 f_{4,X_3 \phi}
+2\dot{\phi}^2f_{4,\phi \phi} 
-4H A_0^2 \left( \dot{A}_0 A_0 f_{4,X_3X_3}
+\dot{\phi} f_{4,X_3 \phi} \right)
+\left[ 2A_0 \left( H \ddot{\phi}+\dot{H} \dot{\phi} 
\right)+\dot{\phi} \left( 2H \dot{A}_0+3H^2 A_0 
\right) \right]f_{5,\phi} \nonumber \\
\hspace{-0.3cm}
& &
-HA_0^2 \left[ 2A_0 \left( \dot{H}+H^2 \right)
+3\dot{A}_0H \right]f_{5,X_3}
+H\dot{\phi}A_0^2 \left( 2\dot{A}_0-HA_0 \right)
f_{5,X_3 \phi}+HA_0 \left( 2\dot{\phi}^2 f_{5,\phi \phi}
-\dot{A}_0 A_0^3 H f_{5,X_3 X_3} \right)=0,\label{back2}\\
\hspace{-0.3cm}
& &
\left( f_{2,X_1}+\dot{\phi}^2 f_{2,X_1 X_1}
+\dot{\phi}A_0 f_{2,X_1X_2}+\frac{1}{4}A_0^2 
f_{2,X_2 X_2} \right) \ddot{\phi}
+3H f_{2,X_1} \dot{\phi}-f_{2,\phi}
+\dot{\phi}^2f_{2,X_1\phi}
-6 \left( \dot{H}+2H^2 \right)f_{4,\phi} \nonumber \\
\hspace{-0.2cm}
& &
+\biggl[ \frac{1}{2}f_{2,X_2}+\frac{1}{2} 
\dot{\phi}^2f_{2,X_1 X_2}+2f_{3,\phi}-3H^2 f_{5,\phi}
+A_0 \left( \dot{\phi}f_{2,X_1 X_3}+\frac{1}{4}
\dot{\phi}f_{2,X_2X_2}-6Hf_{4,X_3 \phi} \right) \nonumber \\
\hspace{-0.3cm}
& &
+\frac{A_0^2}{2} \left( f_{2,X_2 X_3}
-4\tilde{f}_{3,\phi}-6H^2f_{5,X_3 \phi} \right) \biggr]
\dot{A}_0
+\biggl[ \frac{1}{2}\dot{\phi}f_{2,X_2 \phi}
+\frac{3}{2}H f_{2,X_2}+6H f_{3,\phi}
-6A_0H^2 f_{4,X_3 \phi} \nonumber \\
\hspace{-0.3cm}
& &
-3H \left( 2\dot{H}+3H^2 \right)f_{5,\phi}
-A_0^2H^3f_{5,X_3 \phi}
\biggr]A_0=0\,,\label{back3}\\
\hspace{-0.3cm}
& &
2 \left( f_{2,X_3}+6H^2f_{4,X_3}-6H \dot{\phi} f_{4,X_3 \phi} 
\right)A_0-2 \left( 6H f_{3,X_3}+6H \tilde{f}_3
+2\dot{\phi}\tilde{f}_{3,\phi}-3H^3 f_{5,X_3}
+3H^2 \dot{\phi}f_{5,X_3 \phi} \right)A_0^2 \nonumber \\
\hspace{-0.3cm}
& &
+12H^2f_{4,X_3X_3} A_0^3 +2H^3 f_{5,X_3X_3} A_0^4
+\left( f_{2,X2}+4f_{3,\phi}-6H^2 f_{5,\phi} \right)
\dot{\phi}=0\,,
\label{back4}
\ea
where $H=\dot{a}/a$ is the Hubble expansion rate.
As we observe in Eqs.~(\ref{back3}) and (\ref{back4}), 
the scalar field $\phi$ and the temporal vector 
component $A_0$ are coupled to each other in a non-trivial way.
{}From Eq.~(\ref{back4}), we find that $A_0$ depends 
not only on $H$ but also on $\phi$ and $\dot{\phi}$.
In GP theories, $A_0$ depends solely 
on $H$ and hence there exists a de Sitter solution 
characterized by constant $A_0$ 
and $H$ \cite{GPcosmo1,GPcosmo2}. 
In SVT theories, this structure is broken by 
the interaction between $\phi$ and $A_0$, which we need to take into account.

If $A_0$ is the dominant source for the background dynamics 
relevant to cosmic acceleration, the nonvanishing time derivative 
$\dot{\phi}$ leads to the deviation from de Sitter 
solutions characterized by constant $A_0$. 
On the other hand, if the energy density of 
$\phi$ dominates over that of $A_0$, the cosmological 
dynamics of $\phi$ is subject to modifications 
by the existence of $A_0$. 
If we apply this scenario to the early Universe,  
the modification induced by $A_0$ affects the dynamics
of inflation and primordial power spectra of perturbations 
generated during inflation.
If the energy densities of $\phi$ and $A_0$ are comparable 
to each other, there is the possibility for realizing 
``multi-field" inflation driven by the two fields,
even though one of them will play the role of an auxiliary field. It would be also possible to apply the 
above scenario to the dynamics of dark energy 
and possibly to dark matter \cite{Heisenberg18}.

\section{Tensor perturbations}
\label{tensorsec}

We derive the second-order action of tensor perturbations 
for the SVT theories given by the action (\ref{action}). 
Let us consider the linearly perturbed line element 
of intrinsic tensor modes:
\be
ds_t^{2}=-dt^{2}+a^{2}(t) \left( \delta_{ij}
+h_{ij} \right)dx^i dx^j\,,
\ee
where the tensor perturbation $h_{ij}$ obeys the 
transverse and traceless conditions 
$\nabla^j h_{ij}=0$ and ${h_i}^i=0$.

\subsection{Second-order action}

Expanding the action (\ref{action}) up to quadratic 
order in $h_{ij}$ and integrating it by parts, 
the second-order action of tensor 
perturbations yields
\be
{\cal S}_t^{(2)}=\int dt d^3 x\,
\frac{a^3 q_t}{8} \delta^{ik} \delta^{jl} 
\left[ \dot{h}_{ij} \dot{h}_{kl}
-\frac{c_t^2}{a^2} (\partial h_{ij})
(\partial h_{kl})  \right]\,,
\label{St2}
\ee
where the symbol $\partial$ represents the spatial 
partial derivative, and
\ba
q_t &=& 2f_4-2A_0^2 f_{4,{X_3}}
+A_0\dot{\phi}f_{5,\phi}-H A_0^3f_{5,{X_3}}\,,
\label{qt}\\
c_t^2 &=& 
\frac{2f_4-A_0 \dot{\phi}f_{5,\phi}
-\dot{A}_0A_0^2 f_{5,X_3}}{2f_4-2A_0^2 f_{4,{X_3}}
+A_0\dot{\phi}f_{5,\phi}-H A_0^3f_{5,{X_3}}}\,.
\label{ct}
\ea
The terms associated with the tensor mass like 
$\delta^{ik} \delta^{jl}h_{ij}h_{kl}$ vanish 
on account of the background Eq.~(\ref{back2}).
The quantity $c_t$ corresponds to the propagation speed of 
gravitational waves on the FLRW background. 
As we will see in Sec.~\ref{spesec}, there are two 
polarized states for tensor perturbations, 
both of which have the same propagation speed $c_t$.
The existence of additional matter minimally 
coupled to gravity does not affect the value 
of $c_t^2$ given above.
The values of $q_t$ and $c_t^2$ derived in generalized 
Proca theories \cite{GPcosmo1,GPcosmo2}
can be recovered by using the correspondence 
$f_4 \to G_4$, $f_5 \to G_5$, $A_0 \to -\phi$, $X_3 \to X$, 
and $\dot{\phi} \to 0$. Under the two conditions 
\be
q_t>0\,,\qquad c_t^2>0\,,
\label{qtcon}
\ee
there are neither ghost nor Laplacian instabilities 
in the tensor sector.

\subsection{Application to the speed of gravity in late-time 
cosmology}

If we apply SVT theories to the late-time cosmology, 
there is a tight bound $-3 \times 10^{-15} \le c_t-1 
\le 7 \times 10^{-16}$ constrained from the
the gravitational wave event GW170817 \cite{GW170817} together with 
the gamma-ray burst GRB 170817A \cite{Goldstein}. 
{}From Eq.~(\ref{ct}), the SVT theories realizing 
the exact value $c_t=1$ need to satisfy the following 
conditions:
\be
f_4(\phi,X_3)=f_4(\phi)\,,\qquad
f_5(\phi,X_3)={\rm constant}\,.
\label{f45con}
\ee
This mean that $f_4$ does not contain the $X_3$ dependence 
and that $f_5$ depends on neither $\phi$ nor $X_3$. 
This property is similar to what happens in 
scalar-tensor theories with the replacement 
$X_3 \to -\nabla_{\mu}\phi \nabla^{\mu} \phi/2$ \cite{GWdark}.
As expected, the couplings 
$f_2, f_3, \tilde{f}_3, f_6, \tilde{f}_6$ and 
$h_{5j}, \tilde{h}_{5j}$ do not modify the tensor propagation 
speed. If one is willing to apply SVT theories to the late-time 
cosmology, one has to bear in mind the restriction of (\ref{f45con}) 
and include the presence of matter fields. 
Otherwise, for applications to inflation, this restriction can be lifted. 
We will mostly consider the second option here and 
do not take into account matter fields.

\subsection{Tensor power spectrum generated during inflation}
\label{spesec}

If we apply SVT theories to inflation, we do not need to impose the 
conditions (\ref{f45con}). The structure of the second-order action 
(\ref{St2}) is of the same form as that derived in 
Refs.~\cite{KYY,DT11} for Horndeski theories, so it is straightforward 
to compute the primordial tensor power spectrum 
generated during inflation. 
We express $h_{ij}$ in terms of 
the Fourier series, as
\be
h_{ij} ({\bm x}, \tau)=\int \frac{d^3k}{(2\pi)^{3/2}}
e^{i {\bm k} \cdot {\bm x}}
\sum_{\lambda=+,\times} 
\left[ h_{\lambda} (k, \tau) a_{\lambda} ({\bm k})+
h_{\lambda}^* (k, \tau) a_{\lambda}^{\dagger} (-{\bm k})\right] e_{ij}^{(\lambda)} ({\bm k})\,,
\label{gamex1}
\ee
where $\tau=\int a^{-1}dt$ is the conformal time, ${\bm k}$ 
is the coming wavenumber, and $\lambda=+, \times$ denote the
two polarization states.
The polarization tensors $e^{(\lambda)}_{ij} (\bm{k})$ 
obey transverse and traceless conditions 
$k^j e^{(\lambda)}_{ij}=\delta^{ij} e^{(\lambda)}_{ij}=0$
together with the normalization
$\delta^{ik}\delta^{jl}e^{(\lambda)}_{ij} (\bm{k})  
e^{*(\lambda')}_{kl} (\bm{k}) 
= \delta_{\lambda \lambda'}$.
The annihilation and creation operators
$a_{\lambda} ({\bm k})$ and $a_{\lambda}^{\dagger} 
({\bm k}')$ satisfy the commutation relation 
$[a_{\lambda} ({\bm k}),a_{\lambda'}^{\dagger} 
({\bm k}')]=\delta_{\lambda \lambda'}\delta^{(3)}
({\bm k}-{\bm k}')$. 
The primordial power spectrum per unit logarithmic  
wavenumber interval is given by 
\be
{\cal P}_h (k,\tau)=\frac{k^3}{2\pi^2} 
\left( \left| h_{+}(k,\tau)\right|^2
+\left| h_{\times}(k,\tau)\right|^2 \right)\,.
\label{Ptdef}
\ee

We introduce a canonically normalized field $v_{\lambda}(k,\tau)$, as
\be
v_{\lambda}(k,\tau)=z\,h_{\lambda}(k,\tau)\,,\qquad 
z=\frac{a}{2}\sqrt{q_t}\,.
\ee
Varying the action (\ref{St2}) with respect to $h_{ij}$, 
each Fourier component obeys 
\be
v_{\lambda}''+\left( c_t^2 k^2 -\frac{z''}{z} 
\right)v_{\lambda}=0\,,
\label{vlam}
\ee
where a prime represents a derivative with respect to $\tau$.
We consider a quasi de Sitter background on which the variations 
of $H, q_t, c_t$ are small such that $|\dot{H}/H^2| \ll 1$, 
$|\dot{q}_t/(Hq_t)| \ll 1$, and $|\dot{c}_t/(Hc_t)| \ll 1$, 
with the relation $\tau \simeq -(aH)^{-1}$.
Then, the leading-order contribution to 
$z''/z$ is given by $2(aH)^2$. 
For the modes deep inside the 
tensor sound horizon ($c_t^2k^2 \gg a^2H^2$), the solution 
corresponding to the Bunch-Davies vacuum 
is given by $v_{\lambda}=e^{-i c_t k\tau}/\sqrt{2c_tk}$.
On using the de Sitter approximation $z''/z \simeq 2\tau^{-2}$, 
the solution to Eq.~(\ref{vlam}) recovering 
$v_{\lambda}=e^{-i c_t k\tau}/\sqrt{2c_tk}$ in the asymptotic 
past is 
\be
v_{\lambda}(k,\tau)=\frac{i+c_t k |\tau|}
{\sqrt{2} (c_tk)^{3/2}|\tau|}e^{-ic_tk \tau}\,.
\ee
Then, the solution to $h_{\lambda}$ long after the tensor sound 
horizon crossing reduces to
$h_{\lambda}(k,0)=i\sqrt{2/q_t}\,H/(c_tk)^{3/2}$.
{}From Eq.~(\ref{Ptdef}), the leading-order 
primordial power spectrum ${\cal P}_t(k) \equiv {\cal P}_h (k,0)$ yields 
\be
{\cal P}_t(k)=\frac{2H^2}{\pi^2 q_tc_t^3}\,.
\label{Ptf}
\ee
Since the perturbations $h_{\lambda}$ are frozen right after the 
tensor sound horizon crossing, it is sufficient to evaluate the value 
(\ref{Ptf}) at the moment $c_tk=aH$. 
In GR, we have $q_t=M_{\rm pl}^2$ and $c_t^2=1$, 
where $M_{\rm pl}$ is the reduced Planck mass, so 
the tensor power spectrum (\ref{Ptf}) reduces to 
${\cal P}_t(k)=2H^2/(\pi^2 M_{\rm pl}^2)$.
This is modified in SVT theories due to the 
changes of $q_t$ and $c_t^2$. 
We note that the next-to-leading order tensor power spectrum 
can be also computed along the line of Refs.~\cite{Stewart,Chen,DT15}.

\section{Vector perturbations}
\label{vectorsec}

For perturbations in the vector sector, we take the perturbed 
line element in the flat gauge:
\be
ds_v^2=-dt^2+2V_i dt dx^i+a^2(t) \delta_{ij} dx^i dx^j\,,
\ee
where $V_i$ is the vector perturbation obeying the transverse 
condition $\nabla^i V_i=0$. 
At linear order in perturbations, the transverse condition 
translates to $\partial^i V_i=0$, where $\partial^i \equiv \partial/\partial x_i$.
The temporal and spatial  components of $A^{\mu}$ 
associated with the intrinsic vector sector are expressed
in the form 
\be
A_0=A_0(t)\,,\qquad
A_i=Z_i(t,x^i)\,,
\ee
where $Z_i$ is the intrinsic vector perturbation satisfying 
$\partial^i  Z_i=0$. 

For the practical computation, we will consider the 
vector components $V_i=(V_1(t,z), V_2(t,z), 0)$ and 
$Z_i=(Z_1(t,z), Z_2(t,z), 0)$, which automatically 
satisfy the transverse conditions mentioned above.
Expanding Eq.~(\ref{action}) up to quadratic order 
in perturbations and using the background Eqs.~(\ref{back1}) 
and (\ref{back4}), the resulting second-order action 
in the vector sector yields 
\be
{\cal S}_v^{(2)}=\int dt d^3 x \sum_{i=1}^2 
\left[ \frac{aq_v}{2} \dot{Z}_i^2-\frac{1}{2a} 
\alpha_1 (\partial Z_i)^2-\frac{a}{2} \alpha_2 Z_i^2
+\frac{1}{2a} \alpha_3 (\partial V_i)(\partial Z_i) 
+\frac{q_t}{4a} (\partial V_i)^2 \right]\,,
\label{Sv}
\ee
where 
\ba
q_v &=& f_{2,F}+2\dot{\phi}^2 f_{2,Y_1}
+2\dot{\phi}A_0 f_{2,Y_2}+2A_0^2f_{2,Y_3}
-4H \left( \dot{\phi} h_{51}+2A_0  \tilde{h}_{51} \right)
+8H^2 \left( f_6+\tilde{f}_6+\dot{\phi}^2 f_{6,X_1}
+A_0^2 \tilde{f}_{6,X_3} \right)\,,\label{qv}\\
\alpha_1 &=& f_{2,F} - 4\dot{A}_0\tilde{h}_{51} 
+8\left( H^2+\dot{H} \right) \left( f_6+\tilde{f}_6 \right)
-2\ddot{\phi}\,h_{51}
+H \biggl[ 2\dot{\phi} \left( \dot{\phi}^2 h_{52}
-h_{51}+4\ddot{\phi}f_{6,X_1} \right)  \nonumber \\
& &
-A_0 \left\{
4\tilde{h}_{51}-2\dot{\phi}^2
\left( h_{54}+2\tilde{h}_{52} \right) 
-8\dot{A}_0 \tilde{f}_{6,X_3}\right\}
+2\dot{\phi}A_0^2(h_{53}+2\tilde{h}_{54})
+4A_0^3 \tilde{h}_{53}
\biggr]\,,\\
\alpha_2 &=&
f_{2,X_3}+4\dot{H}f_{4,X_3}
-2 \left( \dot{A}_0+3HA_0 \right) 
\left( \tilde{f}_3+f_{3,X_3} \right)
-2\dot{\phi}A_0 \tilde{f}_{3,\phi}
+2H ( 3H f_{4,X_3}+
3HA_0^2 f_{4,X_3 X_3}+2A_0 \dot{A}_0
f_{4,X_3X_3} \nonumber \\
& &-\dot{\phi}f_{4,X_3 \phi})
+H \left( H \dot{A}_0+2 \dot{H} A_0+3H^2 A_0
\right) f_{5,X_3}+H^2 A_0 \left( 
H A_0^2 f_{5,X_3 X_3}+A_0 \dot{A}_0 
f_{5,X_3 X_3}-2\dot{\phi}f_{5,X_3 \phi}\right)\,,\\
\alpha_3 &=& 
-2A_0 f_{4,X_3}-HA_0^2 f_{5,X_3}+\dot{\phi}f_{5,\phi}\,,
\ea
where $q_t$ is defined by Eq.~(\ref{qt}).

Varying the second-order action (\ref{Sv}) with respect to 
$V_i$, it follows that 
\be
\partial^2 \left( \alpha_3 Z_i+q_t V_i \right)=0\,.
\label{ZiVi}
\ee
This is integrated to give $V_i=-\alpha_3 Z_i/q_t$, 
where we set the integration constant 0. 
Substituting this relation into Eq.~(\ref{Sv}), we obtain
\be
{\cal S}_v^{(2)}=\int dt d^3 x \sum_{i=1}^2 
\frac{aq_v}{2} \left[ \dot{Z}_i^2-\frac{c_v^2}{a^2} 
\left( \partial Z_i \right)^2-\frac{\alpha_2}{q_v}Z_i^2 
\right]\,,
\label{Sv2}
\ee
where 
\be
c_v^2=\frac{2\alpha_1 q_t+\alpha_3^2}{2q_t q_v}\,.
\label{cv}
\ee
This result shows that there are two dynamical fields 
$Z_1$ and $Z_2$ in the vector sector with the same 
propagation speed $c_v$. 
Varying the action (\ref{Sv2}) with respect to $Z_i$, 
the equation of motion of vector perturbations 
in Fourier space is given by 
\be
\ddot{Z}_i+\left( H +\frac{\dot{q}_v}{q_v} \right) 
\dot{Z}_i+\left( c_v^2\frac{k^2}{a^2}+
\frac{\alpha_2}{q_v} \right) Z_i=0\,,
\ee
where the term $m_v^2=\alpha_2/q_v$ corresponds to the 
vector mass squared. In the limit that 
$c_v^2 k^2/a^2 \gg m_v^2$, the mass term 
is irrelevant to the dynamics of vector perturbations, 
so there are neither ghost nor 
Laplacian instabilities under the conditions 
\be
q_v>0\,,\qquad c_v^2>0\,.
\label{qvcvcon}
\ee

{}From Eq.~(\ref{qv}), we find that the functional 
dependence $f_2(F,Y_1,Y_2, Y_3)$ 
as well as the functions 
$h_{51}, \tilde{h}_{51}, f_6, \tilde{f}_6$ themselves 
affect the no-ghost condition of vector perturbations. 
The value of $q_v$ derived in GP 
theories \cite{GPcosmo2} can be recovered by 
using the correspondence 
$\dot{\phi} \to 0$, $X_3 \to X$, 
$Y_3 \to Y$, $A_0 \to -\phi$, $f_2 \to G_2$, 
$\tilde{h}_{51} \to -g_5(X)/2$, 
$f_6 \to 0$, and $4\tilde{f}_6 \to G_6(X)$ 
in Eq.~(\ref{qv}).
Since Eq.~(\ref{cv}) contains $q_t, \alpha_1, \alpha_3$, 
the vector propagation speed is also affected by the dependence 
of $f_4(X_3)$, $f_5(\phi,X_3)$ and the functions 
$h_{52}, \tilde{h}_{52}, h_{53}, \tilde{h}_{53}, 
h_{54}, \tilde{h}_{54}$.

If we apply SVT theories to inflation, the evolution of 
$Z_i$ is different depending on the mass term $m_v^2$. 
If the condition $m_v^2 \gg H^2$ is satisfied during inflation, 
the vector perturbation is subject to strong suppression for 
the modes $c_v^2 k^2/a^2<m_v^2$ (analogous to what happens for tensor perturbations in a Lorentz-violating 
massive gravity scenario studied in Ref.~\cite{Kuro}). 
If $m_v^2 \ll H^2$, then the vector perturbation is not 
suppressed even after the vector horizon crossing 
($c_v^2 k^2/a^2<H^2$). 
We leave the detailed analysis for the computation of 
the primordial vector power spectrum generated during 
inflation as a future work.

\section{Scalar perturbations}
\label{scalarsec}

For scalar perturbations, we consider the linearly perturbed 
line-element in the flat gauge:
\be
ds_s^{2}=-(1+2\alpha)\,dt^{2}+2\partial_{i}\chi dt\,dx^{i}
+a^{2}(t) \delta_{ij}dx^i dx^j\,, 
\ee
where $\alpha$ and $\chi$ are scalar metric perturbations. 
We write the components of the vector field in the form 
\be
A^{0}=-A_0(t)+\delta A\,,\qquad 
A_{i}=\partial_i\psi\,,
\ee
where $\delta A$ is the perturbation of the temporal vector 
component $A^0$, and $\psi$ is the longitudinal scalar perturbation.
The scalar field $\phi$ is decomposed into the background 
and perturbed parts, as 
\be
\phi=\phi_0(t)+\delta \phi\,,
\ee
where, in the following, we omit the subscript ``0'' from the background 
value of $\phi$.

We expand the action (\ref{action}) up to quadratic order in scalar perturbations 
$\alpha, \chi, \delta A, \psi, \delta \phi$. 
In doing so, we use the background Eqs.~(\ref{back1}) and (\ref{back4}) to 
eliminate the terms $f_2$ and $f_{3,\phi}$. 
Then, the second-order action for scalar perturbations can be expressed as 
\be
{\cal S}_s^{(2)}=\int dt d^3x \left({\cal L}_s^{\phi}+{\cal L}_s^{\rm GP}\right)\,, 
\label{Ss}
\ee
where 
\ba
&&
{\cal L}^{\phi}_s=a^3\left[
D_1\dot{\dphi}^2+D_2\frac{(\partial\dphi)^2}{a^2}+D_3\dphi^2
+\left(D_4\dot{\dphi}+D_5\dphi+D_6\frac{\partial^2\dphi}{a^2}\right) \alpha
-\left(D_6\dot{\dphi}-D_7\dphi\right)\frac{\partial^2\chi}{a^2}
\right.\notag\\
&&
\left.\hspace{1.5cm}
+\left(D_8\dot{\dphi}+D_9\dphi\right)\da+D_{10}\,\dphi\,\frac{\partial^2\psi}{a^2}
\right]\,, 
\label{Lphi}
\ea
and 
\ba
&&
{\cal L}^{\rm GP}_s=a^3\left[
\left(w_1\alpha-w_2\frac{\da}{A_0}\right)\frac{\partial^2\chi}{a^2}
-w_3\frac{(\partial\alpha)^2}{a^2}+w_4\alpha^2
-\left(w_3\frac{\partial^2\da}{a^2A_0}-w_8\frac{\da}{A_0}
+w_3\frac{\partial^2\dot{\psi}}{a^2A_0}+w_6\frac{\partial^2\psi}{a^2}\right)\alpha
\right.\notag\\
&&\hspace{1.75cm}\left.
-w_3\frac{(\partial\da)^2}{4a^2A_0^2}+w_5\frac{\da^2}{A_0^2}
+\left\{w_3\dot{\psi}-(w_2-A_0w_6)\psi\right\}\frac{\partial^2\da}{2a^2A_0^2}
-w_3\frac{(\partial\dot{\psi})^2}{4a^2A_0^2}+w_7\frac{(\partial\psi)^2}{2a^2}
\right]\,.
\label{LGP} 
\ea
The coefficients $D_{1,\cdots,10}$ and $w_{1,\cdots,8}$ are given 
in Appendix. The Lagrangian ${\cal L}_s^{\phi}$ arises from the scalar perturbation $\delta \phi$. 
The other Lagrangian ${\cal L}_s^{\rm GP}$ has the similar
structure to that in GP theories \cite{GPcosmo1}. 
In GP theories, the coefficient $w_8$ is related to $w_1$ and $w_4$, as 
$w_8=3Hw_1-2w_4$ \cite{GPcosmo1}, while, in the present case,
this relation is modified to  $w_8=3Hw_1-2w_4-\tp D_4$ by the presence 
of the scalar field $\phi$. The effect of intrinsic vector modes on scalar 
perturbations appears only through $w_3=-2A_0^2q_v$, 
where $q_v$ is given by Eq.~(\ref{qv}).

Since there are no time derivatives of $\alpha,\chi,\da$ in Eqs.~(\ref{Lphi}) 
and (\ref{LGP}), these fields are non-dynamical. 
On the other hand, the perturbations $\psi$ and $\dphi$ correspond to the 
dynamical fields in the scalar sector. 
The field $\psi$ is the longitudinal scalar component of vector field 
associated with the breaking of $U(1)$ gauge symmetry, whereas
the perturbation $\dphi$ arises from the scalar field $\phi$.
Varying the action (\ref{Ss}) with respect to $\alpha,\chi,\da$, we 
obtain the three constraint equations in Fourier space: 
\ba
&&
D_4\dot{\dphi}+D_5\dphi+2w_4\alpha+w_8\frac{\da}{A_0}
+\frac{k^2}{a^2}\left(w_3\frac{\dot{\psi}}{A_0}+w_6\psi-D_6\dphi
-2w_3\alpha-w_1\chi+w_3\frac{\da}{A_0}\right)=0\,,
\label{eqalpha}\\
&&
D_6\dot{\dphi}-D_7\dphi-w_1\alpha+w_2\frac{\da}{A_0}=0\,,
\label{eqchi}\\
&&
D_8\dot{\dphi}+D_9\dphi+w_8\frac{\alpha}{A_0}+2w_5\frac{\da}{A_0^2}
-\frac{k^2}{a^2}\frac{1}{A_0} 
\left(\frac{w_3}{2}\frac{\dot{\psi}}{A_0}
+\frac{A_0w_6-w_2}{2}\frac{\psi}{A_0}-w_3\alpha-w_2\chi
+\frac{w_3}{2}\frac{\da}{A_0}\right)=0\,. 
\label{eqdA}
\ea
We solve Eqs.~(\ref{eqalpha})-(\ref{eqdA}) for 
$\alpha,\chi,\da$ and eliminate these variables from 
the action (\ref{Ss}). 
Then, the second-order action of scalar perturbations 
can be expressed in the form 
\be
{\cal S}_s^{(2)}=\int dt d^3x\,a^{3}\left( 
\dot{\vec{\mathcal{X}}}^{t}{\bm K}\dot{\vec{\mathcal{X}}}
-\frac{k^2}{a^2}\vec{\mathcal{X}}^{t}{\bm G}\vec{\mathcal{X}}
-\vec{\mathcal{X}}^{t}{\bm M}\vec{\mathcal{X}}
-\vec{\mathcal{X}}^{t}{\bm B}\dot{\vec{\mathcal{X}}}
\right)\,,
\label{Ss2}
\ee
where ${\bm K}$, ${\bm G}$, ${\bm M}$, ${\bm B}$ 
are $2 \times 2$ matrices, and $\vec{\mathcal{X}}$ 
is defined by
\be
\vec{\mathcal{X}}^{t}=\left(\psi, \dphi \right) \,.
\ee
In the small-scale limit, the leading-order contributions 
to the matrix ${\bm M}$ do not contain the $k^2$ terms. 
We shift the $k^2$ terms appearing in ${\bm B}$ to 
the matrix components of  ${\bm G}$ 
after integrating them by parts. Then, in the $k \to \infty$ limit, 
the components of ${\bm K}$ and 
${\bm G}$ are given, respectively, by 
\ba
&&
K_{11}=\frac{w_1^2w_5+w_2^2w_4+w_1w_2w_8}{A_0^2(w_1-2w_2)^2}\,,\notag\\
&&
K_{22}=D_1+\frac{D_6}{w_1-2w_2}\left(D_4
+\frac{w_4+4w_5+2w_8}{w_1-2w_2}D_6+2A_0D_8\right)
\,,\notag\\
&&
K_{12}=K_{21}=-\frac{1}{2A_0(w_1-2w_2)}\left[w_2D_4
+\frac{w_1(4w_5+w_8)+2w_2(w_4+w_8)}{w_1-2w_2}D_6
+A_0w_1D_8\right]\,,
\label{Kmat}
\ea
and 
\ba
&&
G_{11}=\dot{E_1}+H E_1-\frac{4A_0^2}{w_3}E_1^2-\frac{w_7}{2}\,,\notag\\
&&
G_{22}=\dot{E_2}+H E_2-\frac{2A_0}{w_2}D_7E_3
-\frac{4A_0^2}{w_3}E_3^2-D_2\,,\notag\\
&&
G_{12}=G_{21}=\dot{E_3}+H E_3-\frac{4A_0^2}{w_3}E_1E_3
+\frac{w_2}{2A_0(w_1-2w_2)}D_7+\frac{D_{10}}{2}\,,
\ea
where we introduced 
\be
E_1=\frac{w_6}{4A_0}-\frac{w_1w_2}{4A_0^2(w_1-2w_2)}\,,\qquad 
E_2=-\frac{D_6^2}{2(w_1-2w_2)}\,,\qquad 
E_3=\frac{w_2D_6}{2A_0(w_1-2w_2)}\,.
\ee

In order to ensure the absence of scalar ghosts, the kinetic matrix 
${\bm K}$ must be positive definite. In other words, the determinants of 
principal sub-matrices of ${\bm K}$ need to be positive. 
Thus, we require the following two no-ghost conditions: 
\ba
&&
K_{11}>0
\quad {\rm or} \quad 
K_{22}>0\,,
\label{ng1}\\
&&
q_s \equiv K_{11}K_{22}-K_{12}^2 >0\,. 
\label{ng2}
\ea

In the small-scale limit, the dispersion relation following from the action (\ref{Ss2}) 
with frequency $\omega$ is given by 
\be
{\rm det}\left(\omega^2{\bm K}-\frac{k^2}{a^2}{\bm G}\right)=0\,.
\ee
Introducing the scalar sound speed 
$c_s$ as $\omega^2=c_s^2k^2/a^2$, 
the above dispersion relation leads to the two scalar propagation speed squares: 
\ba
\hspace{-0.9cm}
c_{s1}^2
&=&
\frac{K_{11}G_{22}+K_{22}G_{11}-2K_{12}G_{12}
+\sqrt{(K_{11}G_{22}+K_{22}G_{11}-2K_{12}G_{12})^2
-4(K_{11}K_{22}-K_{12}^2)(G_{11}G_{22}-G_{12}^2)}}
{2(K_{11}K_{22}-K_{12}^2)},
\label{cs1}\\
\hspace{-0.9cm}
c_{s2}^2
&=&
\frac{K_{11}G_{22}+K_{22}G_{11}-2K_{12}G_{12}
-\sqrt{(K_{11}G_{22}+K_{22}G_{11}-2K_{12}G_{12})^2
-4(K_{11}K_{22}-K_{12}^2)(G_{11}G_{22}-G_{12}^2)}}
{2(K_{11}K_{22}-K_{12}^2)}.
\label{cs2}
\ea
For the absence of Laplacian instabilities, 
we require that 
\be
c_{s1}^2 > 0\,,\qquad 
c_{s2}^2 > 0\,.
\ee

{}From Eqs.~(\ref{ng1})-(\ref{ng2}), the functions 
$f_3,\tilde{f}_3,f_4,f_5$ as well as the functional 
dependence of $X_1, X_2, X_3$ in $f_2$ affect 
no-ghost conditions of scalar perturbations.
Since the matrix components $G_{11}, G_{22}, G_{12}$ contain 
the term $w_3=-2A_0^2q_v$, the propagation speeds $c_{s1}$ 
and $c_{s2}$ are affected by intrinsic vector modes. 

If we apply SVT theories to ``multi-field" inflation 
driven by the background field $\phi(t)$ and the auxiliary field $A_0(t)$, both $\delta \phi$ 
and $\psi$ contribute to the curvature perturbation 
${\cal R}$ \cite{Bassett}. 
In such cases, the separation between adiabatic and isocurvature 
perturbations is useful for the computation of 
primordial scalar power spectrum generated during inflation \cite{Gordon}. 
The resulting power spectra of curvature and isocurvature perturbations 
as well as their correlations depend on the models of inflation \cite{corre}, so 
the observational prediction of SVT theories relevant to CMB 
temperature anisotropies
deserves for separate detailed analysis in future.

\section{Application to a concrete model of inflation}
\label{apsec}

Let us apply the $U(1)$ broken SVT theories to the background 
dynamics of inflation and reheating as well as to the stability 
conditions in these epochs. We consider the following model 
\ba
f_2 &=& F+X_1-V(\phi)+\beta_{m} M X_2+\beta_{A} M^2 X_3\,,
\nonumber \\
f_4 &=& \frac{M_{\rm pl}^2}{2}+\beta_{G}X_3\,,
\label{model}
\ea
where $V(\phi)$ is a scalar potential, $M$ is a constant having 
a dimension of mass (of order the Hubble expansion rate during inflation),
and $\beta_m, \beta_A, \beta_{G}$ are 
dimensionless coupling constants.
The other functions $f_3, \tilde{f}_3, f_5, h_{5j}, \tilde{h}_{5j}, f_6, \tilde{f}_6$ 
in the action (\ref{action}) are taken to be 0. 

\subsection{Background dynamics}

For the above model, the background Eqs.~(\ref{back1})-(\ref{back4}) reduce, respectively, to
\ba
& &
3M_{\rm pl}^2 H^2=\frac{1}{2} \dot{\phi}^2+V
-\frac{1}{2}\beta_A M^2 A_0^2+3\beta_{G}H^2A_0^2 \,,
\label{back1d}\\
& &
\left(  2\dot{H}+3H^2 \right)
\left(  M_{\rm pl}^2-\beta_G A_0^2 \right)
=-\frac{1}{2}\dot{\phi}^2+V
-\frac{1}{2}\beta_m M \dot{\phi}A_0
-\frac{1}{2}\beta_A M^2 A_0^2
+4\beta_G H \dot{A}_0 A_0\,,
\label{back2d}\\
& &
\ddot{\phi}+3H \dot{\phi}+V_{,\phi}+
\frac{1}{2}M \beta_m \left( \dot{A}_0
+3H A_0 \right)=0\,,
\label{back3d}\\
& &
A_0=-\frac{\beta_m M}
{2(\beta_A M^2+6\beta_G H^2)} \dot{\phi}\,.
\label{back4d}
\ea
As we see in Eqs.~(\ref{back3d}) and (\ref{back4d}), the 
nonvanishing coupling $\beta_m$ induces a mixing between 
the scalar derivative $\dot{\phi}$ and the temporal vector component $A_0$. 
If $\beta_m=0$, then the system reduces to  
the single-field slow-roll inflation with $A_0=0$.
During the inflationary stage in which $H$ is nearly constant, 
the ratio $A_0/\dot{\phi}$ stays nearly constant. 
This property also holds for the case in which the condition 
$|\beta_A M^2| \gg |6\beta_G H^2|$ is satisfied.

To study the dynamics of inflation, it is convenient to define the 
following slow-roll parameters \cite{Bassett}:
\be
\epsilon \equiv -\frac{\dot{H}}{H^2}\,,\qquad 
\epsilon_V \equiv \frac{M_{\rm pl}^2}{2} 
\left( \frac{V_{,\phi}}{V} \right)^2\,, \qquad 
\eta \equiv \frac{\ddot{\phi}}{H \dot{\phi}}\,,
\label{slowroll}
\ee
whose orders are less than 1 during inflation.
Substituting Eq.~(\ref{back4d}) and its time derivative 
into Eq.~(\ref{back2d}) and using Eq.~(\ref{back1d}) to eliminate $V$, 
we find that $\epsilon$ can be expressed as 
\be
\epsilon=\frac{\dot{\phi}^2 
(\beta_A M^2+6\beta_G H^2)
[144\beta_G^2 H^4+2\beta_G H^2 M^2 
\{ 24\beta_A-(3+2\eta)\beta_m^2 \}
+\beta_AM^4 (4\beta_A-\beta_m^2)]}
{2H^2 [4M_{\rm pl}^2(\beta_A M^2+6\beta_G H^2)^3
+\beta_m^2 \beta_GM^2 (18\beta_G H^2-\beta_A M^2)\dot{\phi}^2]}\,.
\label{ep}
\ee

We employ the approximation that 
the three slow-roll parameters defined in Eq.~(\ref{slowroll}) are smaller than the order 1.
{}From Eqs.~(\ref{back3d}) and (\ref{back4d}), 
it follows that 
\be
\dot{\phi} \simeq -\frac{4V_{,\phi} 
(\beta_A M^2+6\beta_G H^2)}
{3H (4\beta_A M^2+24\beta_G H^2
-\beta_m^2 M^2)}\,.
\label{dotphi}
\ee
Substituting Eq.~(\ref{back4d}) into Eq.~(\ref{back1d}), 
the ratio between the vector kinetic energy 
$K_A=-\beta_A M^2 A_0^2/2+3\beta_G H^2 A_0^2$ 
and the scalar kinetic energy $K_{\phi}=\dot{\phi}^2/2$ 
is given by 
\be
\frac{K_A}{K_{\phi}}
=\frac{M^2 (6\beta_G H^2-\beta_A M^2)}
{4(6\beta_G H^2+\beta_A M^2)^2}\beta_m^2\,.
\label{KAp}
\ee
In the following,  we assume that $\beta_m^2$ is 
at most of the orders $|\beta_G|$ and $|\beta_A|$, 
in which case $K_A$ does not exceed $K_{\phi}$.
Then, the scalar potential $V$ dominates over 
the other terms on the 
r.h.s. of Eq.~(\ref{back1d}) during slow-roll inflation, so 
the Hubble expansion rate can be estimated as $H \simeq \sqrt{V/3}/M_{\rm pl}$. 
We substitute this relation and Eq.~(\ref{dotphi}) into Eq.~(\ref{ep}) and 
neglect the slow-roll parameters $\eta$ and $\epsilon_V$ relative to 1 in the end. 
This process leads to 
\be
\epsilon \simeq \epsilon_V 
\left[ 1+\frac{\beta_m^2 M^2M_{\rm pl}^2}
{4(\beta_A M^2M_{\rm pl}^2+2\beta_G V)} \right]\,,
\label{epes}
\ee
where we employed the approximation that $|\beta_m| \ll 1$ 
and picked up the leading-order contribution 
of $\beta_m$ to $\epsilon$.
For $\beta_m=0$, we have $\epsilon \simeq \epsilon_V$ 
as in standard slow-roll inflation, but the presence of the 
coupling $\beta_m$ leads to $\epsilon \neq \epsilon_V$.
If $\beta_A M^2M_{\rm pl}^2+2\beta_G V>0$, 
then $\epsilon$ is larger than $\epsilon_V$.
In this case, the mixing between $\dot{\phi}$ and $A_0$ 
effectively leads to the faster evolution of inflaton, 
so inflation tends to be less efficient for a given potential $V(\phi)$.

\begin{figure}[h]
\begin{center}
\includegraphics[height=3.4in,width=3.4in]{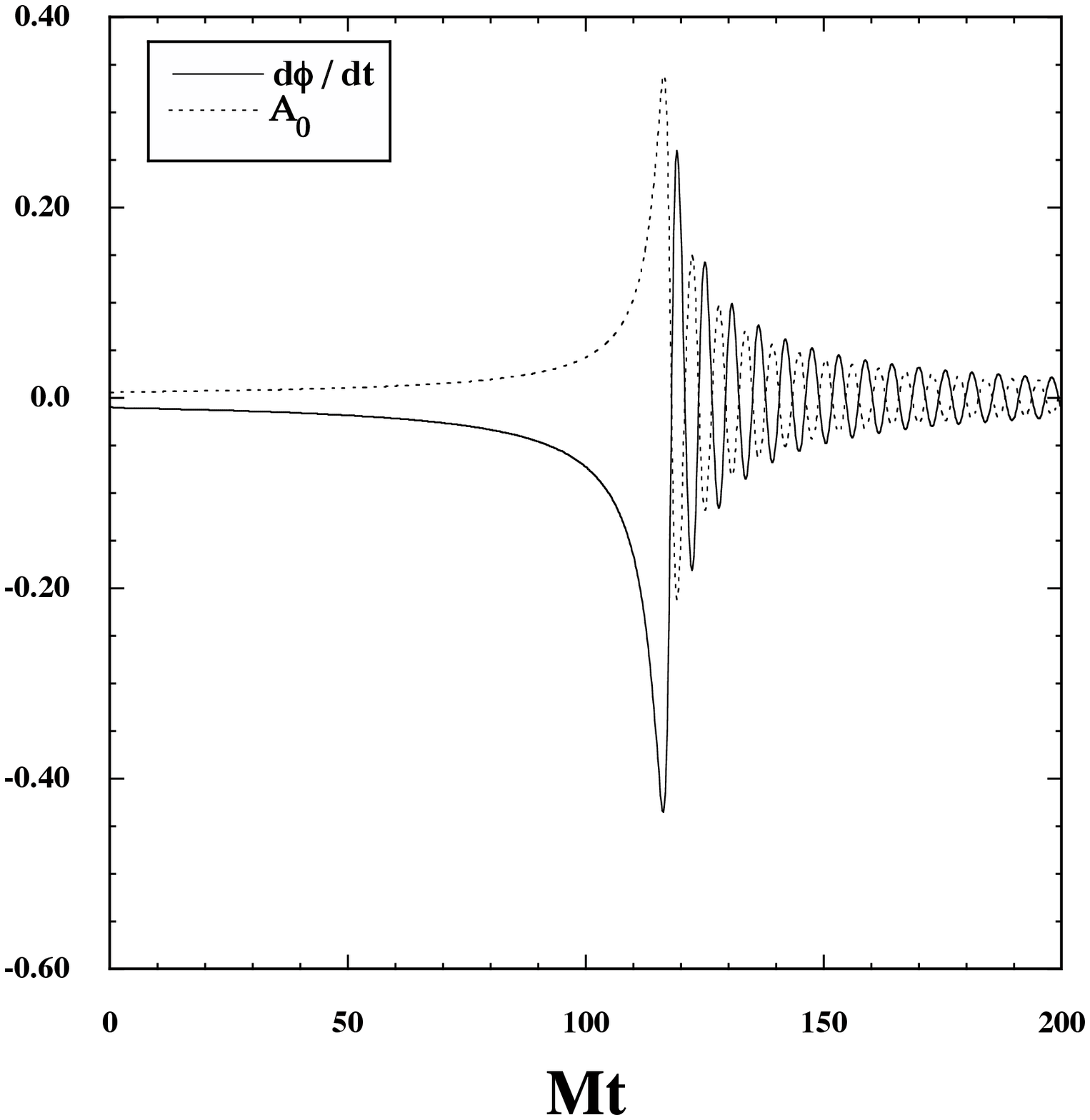}
\includegraphics[height=3.4in,width=3.6in]{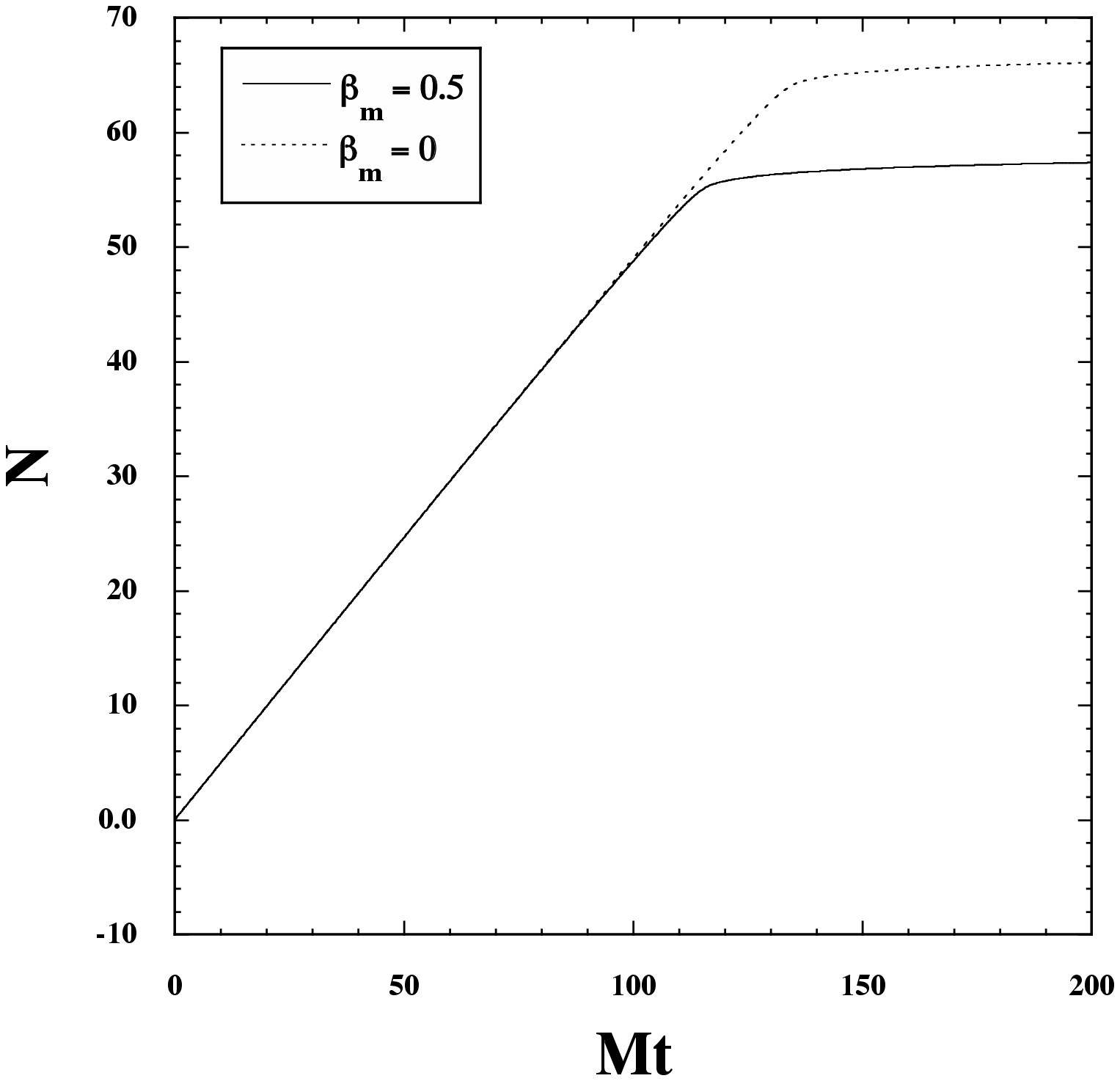}
\end{center}
\caption{\label{fig1}
(Left) Evolution of $\dot{\phi}$ and $A_0$ (normalized by 
$MM_{\rm pl}$ and $M_{\rm pl}$ respectively) during 
inflation and reheating for the potential (\ref{potential}) 
with $\alpha_c=\sqrt{6}/3$. 
We choose the coupling constants 
$\beta_m=0.5$, $\beta_A=0.3$, and 
$\beta_{G}=0.1$ with the initial conditions 
$\phi=5.5M_{\rm pl}$ and $\dot{\phi}=-9.0 \times 10^{-3}MM_{\rm pl}$. 
For the consistency with Eqs.~(\ref{back1d}) and (\ref{back4d}), 
the initial values of $H$ and $A_0$ are chosen to be
$H=0.4944M$ and $A_0=5.0374 \times 10^{-3}M_{\rm pl}$.
(Right) Evolution of the number of e-foldings $N$ for 
the case shown in the left panel (plotted as a solid line). 
The dotted line corresponds to the case 
$\beta_m=0$ and $\alpha_c=\sqrt{6}/3$ 
(i.e., standard Starobinsky inflation) 
with the same initial conditions of $\phi$ and $\dot{\phi}$ 
as those used in the left panel.
}
\end{figure}

To confirm the above analytic estimation, we consider the $\alpha$-attractor 
model given by the potential \cite{Kallosh}
\be
V(\phi)=\frac{M^2 M_{\rm pl}^2}{2\alpha_c^2} 
\left( 1-e^{-\alpha_c \phi/M_{\rm pl}} \right)^2\,,
\label{potential}
\ee
where $\alpha_c$ is a positive constant.  
The Starobinsky inflation \cite{Starobinsky} corresponds 
to $\alpha_c=\sqrt{6}/3$ in the Einstein frame \cite{fR}. 
Inflation occurs for $\alpha_c \phi/M_{\rm pl} \gg 1$, 
in which regime the potential is nearly constant: 
$V(\phi) \simeq M^2M_{\rm pl}^2/(2\alpha_c^2)$. 
Then, the Hubble expansion rate during inflation $H_{\rm inf}$ 
is related to the mass $M$, as $H_{\rm inf} \simeq M/(\sqrt{6}\alpha_c)$. 
For $\alpha_c={\cal O}(1)$, $H_{\rm inf}$ is of the same order as $M$.
The system enters the reheating 
stage for $|\alpha_c \phi/M_{\rm pl}| \lesssim 1$, during which 
the potential is approximately given by $V(\phi) \simeq M^2 \phi^2/2$.

In the left panel of Fig.~\ref{fig1}, we show the numerical 
solutions to $\dot{\phi}$ and $A_0$ versus $Mt$ 
for the Starobinsky potential ($\alpha_c=\sqrt{6}/3$) with 
the couplings $\beta_m=0.5$, $\beta_A=0.3$, and $\beta_G=0.1$. 
The initial conditions of $\phi$, $\dot{\phi}$, and $A_0$ are chosen to be consistent with Eqs.~(\ref{back1d}) and (\ref{back4d}). 
As estimated from Eq.~(\ref{back4d}), the ratio between $A_0$ and $\dot{\phi}$ 
stays nearly constant ($A_0/M_{\rm pl} \simeq 
-0.56\,\dot{\phi}/(MM_{\rm pl}$)) during inflation. 
In the reheating stage, the term $6 \beta_G H^2$ starts to be 
negligible relative to $\beta_A M^2$ 
due to the decrease of $H$. 
Then, the amplitude of $A_0$ decreases in the same way as that of 
$\dot{\phi}$ according to the relation $A_0/M_{\rm pl}
=-[\beta_m/(2\beta_A)]\,\dot{\phi}/(MM_{\rm pl})=-0.83\,\dot{\phi}/(MM_{\rm pl})$. 
In the left panel of Fig.~\ref{fig1}, we can confirm that $\dot{\phi}$ 
and $A_0$ slowly evolve during inflation with 
the relation mentioned above and that they oscillate during reheating with the asymptotic behavior 
$A_0/\dot{\phi}={\rm constant}$.

In the right panel of Fig.~\ref{fig1}, the number 
of e-foldings $N=\ln a$ from the onset of inflation 
is plotted for the model parameters and initial conditions 
same as those used in the left panel.
We also show the evolution of $N$ for 
$\beta_m=0$, i.e., Starobinsky inflation with $A_0=0$.
For $\beta_m=0.5$, the value of $N$ reached at the end of inflation 
is smaller than that for $\beta_m=0$ about by 14 \%. 
This is consistent with the fact that 
the slow-roll parameter (\ref{epes}) for $\beta_m=0.5$ 
can be estimated as $\epsilon \simeq 1.14 \epsilon_V$. 
Thus, the nonvanishing coupling $\beta_m$ leads to 
the smaller amount of inflation  due to the additional 
evolution of $A_0$ besides $\dot{\phi}$.

\subsection{Stability conditions}

For the model (\ref{model}), the quantities $q_t$ and $c_t^2$ 
are given, respectively, by 
\be
q_t=M_{\rm pl}^2-\frac{\beta_m^2 \beta_G  M^2 
\dot{\phi}^2}{4(\beta_A M^2+6\beta_G H^2 )^2}\,,\qquad
c_t^2=1+\frac{2\beta_m^2 \beta_G M^2 
\dot{\phi}^2}{4(\beta_A M^2+6\beta_G H^2 )^2M_{\rm pl}^2
-\beta_m^2\beta_G M^2 
\dot{\phi}^2}\,,
\label{qct}
\ee
where we used Eq.~(\ref{back4d}) to express $A_0$ 
in terms of $\dot{\phi}$.
Then, the stability conditions (\ref{qtcon}) of tensor perturbations translate to 
\be
-4 \left(\beta_A M^2+6\beta_G H^2 \right)^2 M_{\rm pl}^2
\le \beta_m^2 \beta_G  M^2 \dot{\phi}^2
< 4 \left(\beta_A M^2+6\beta_G H^2 \right)^2 M_{\rm pl}^2\,.
\label{tencon}
\ee
For vector perturbations, we have 
\be
q_v=1\,,\qquad 
c_v^2=1+\frac{2\beta_m^2 \beta_G^2 M^2 \dot{\phi}^2}
{4(\beta_A M^2+6\beta_G H^2 )^2M_{\rm pl}^2
-\beta_m^2\beta_G M^2 
\dot{\phi}^2}\,,
\label{qvt}
\ee
so the no-ghost condition is automatically satisfied.
Under the conditions (\ref{tencon}),
there are no Laplacian instabilities of vector perturbations.

For scalar perturbations, the quantity $K_{22}$ defined in Eq.~(\ref{Kmat}) is given by 
\be
K_{22}=\frac{1}{2}\,,
\label{K22}
\ee
and hence the latter condition of Eq.~(\ref{ng1}) is always 
satisfied. The other no-ghost condition translates to 
\ba
q_s&=&
\left[\bem^2\beG M^2\left\{(\bem^2-4\beA)M^2+72\beG H^2\right\}\tp^2
-4\Mpl^2(\beA M^2+6\beG H^2)^2\left\{(\bem^2-4\beA)M^2-24 \beG H^2\right\}\right]
\notag\\
&&
\times\left[4(6\beG H^2+\beA M^2)^2 \Mpl^2-\bem^2 \beG M^2 \tp^2\right]
/\left[16\left\{4(6\beG H^2+\beA M^2)^2 \Mpl^2-3 \bem^2 \beG M^2 \tp^2\right\}^2\right]
>0\,.
\label{scacon}
\ea
If $\beta_G=0$, then the condition (\ref{scacon}) 
translates to $\beta_A>\beta_m^2/4$.
To satisfy this inequality, it is necessary to have 
$\beta_A>0$, which means that the mass squared 
of the vector field $A_{\mu}$ is positive.

\begin{figure}[h]
\begin{center}
\includegraphics[height=3.4in,width=3.4in]{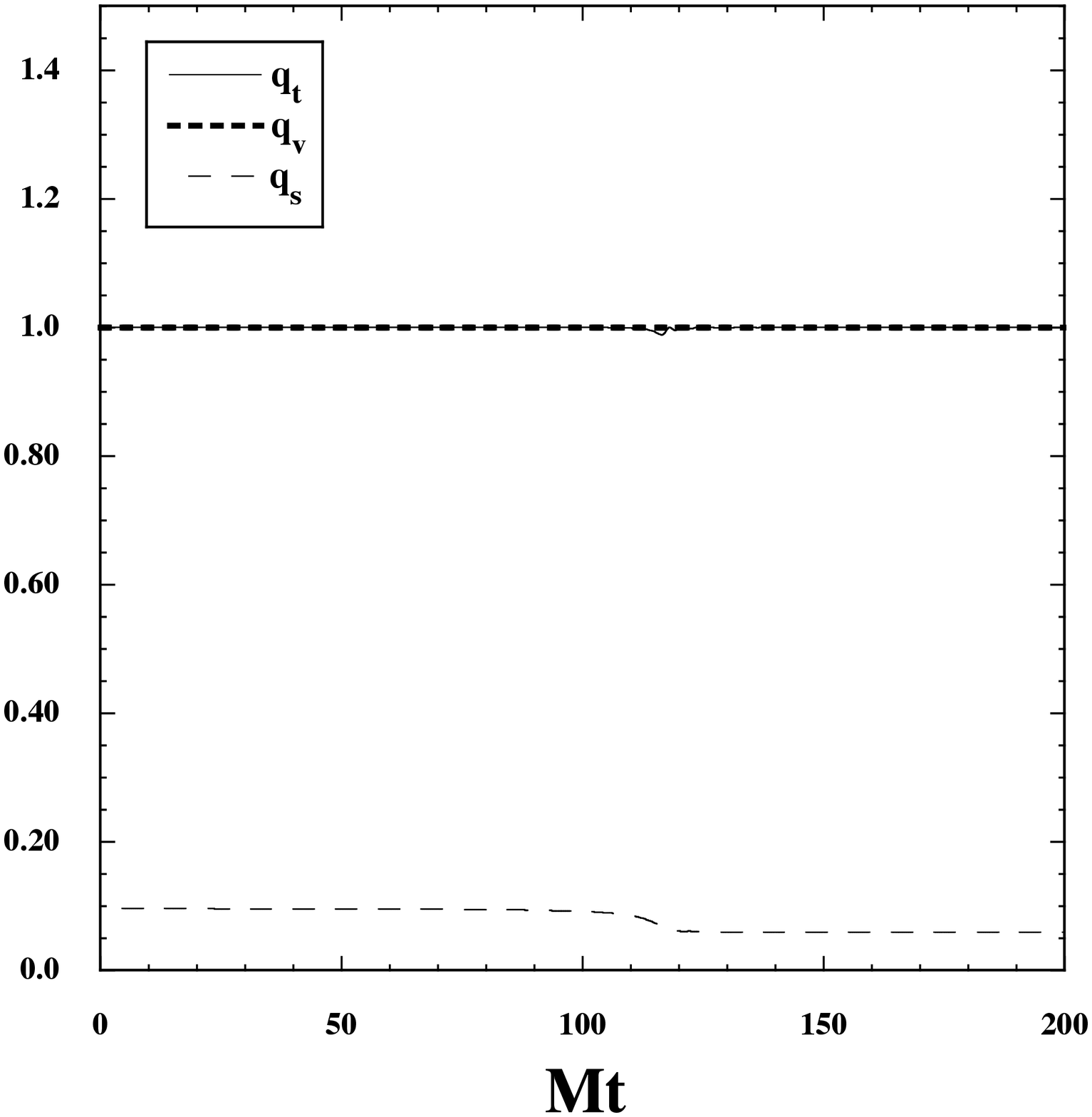}
\includegraphics[height=3.4in,width=3.4in]{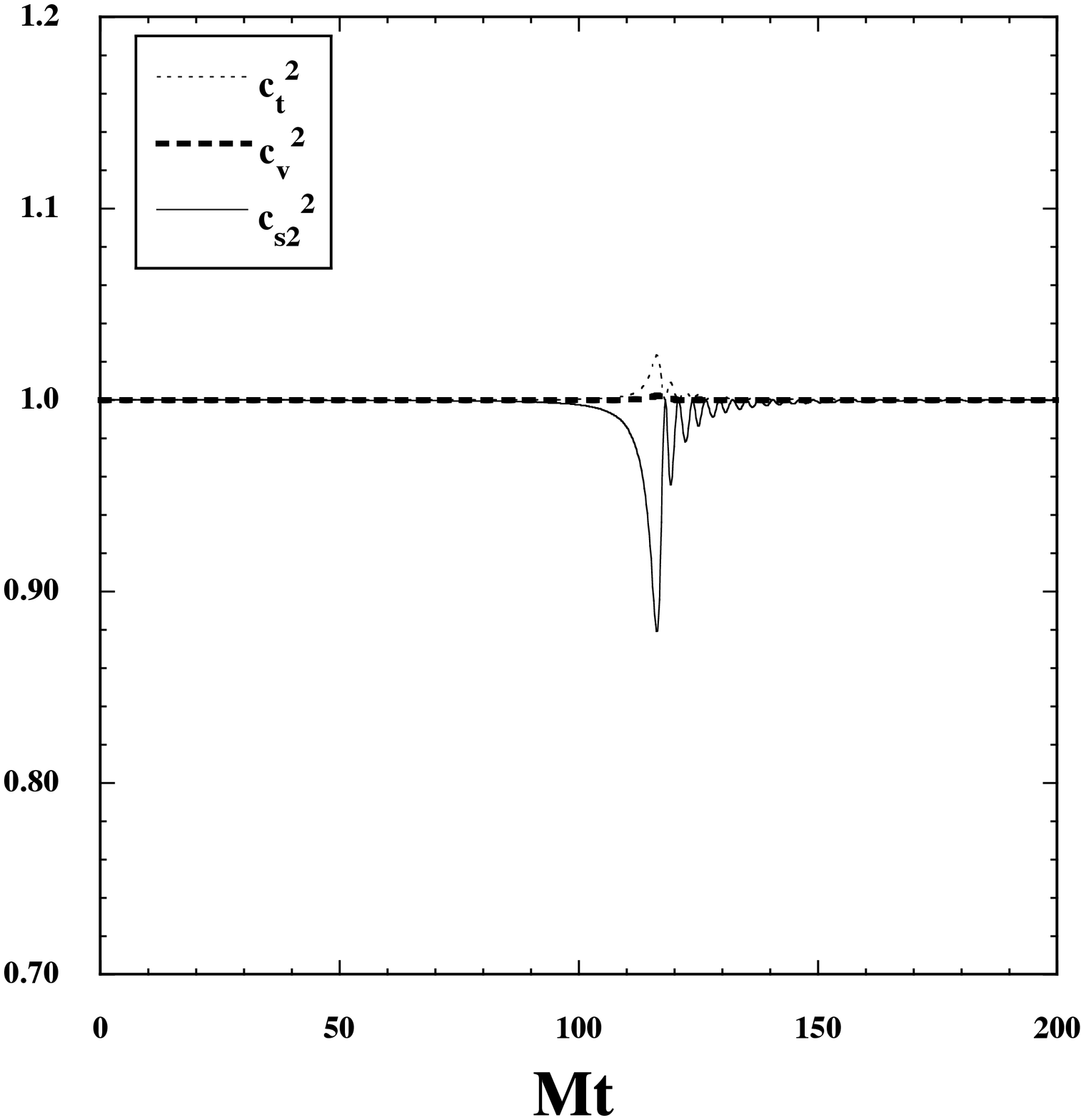}
\end{center}
\caption{\label{fig2}
Evolution of $q_t, q_v, q_s$ (left) and 
$c_{t}^2, c_{v}^2, c_{s2}^2$ (right) for the same model parameters and initial conditions 
as those used in the left panel of Fig.~\ref{fig1}. 
Note that $q_t$ and $q_s$ are normalized by 
$M_{\rm pl}^2$ and $M^2$, respectively.}
\end{figure}

In addition to Eq.~(\ref{K22}), we also have 
$G_{22}=1/2$ and $K_{12}=G_{12}$ 
for the model (\ref{model}). Substituting them into 
Eqs.~(\ref{cs1})-(\ref{cs2}), it follows that one of 
the scalar propagation speed squares reduces to 
\be
c_{s1}^2=1\,,
\ee
while the other is given by  
$c_{s2}^2=(G_{11}-2K_{12}^2)/(K_{11}-2K_{12}^2)
=1+(G_{11}-K_{11})/(2q_s)$. 
More explicitly, the latter is expressed as 
\ba
c_{s2}^2
&=&
1+\frac{1}{2q_s(\Mpl^2-3\beG A_0^2)^2}\left[
2\beG(\Mpl^2-3\beG A_0^2)(\Mpl^2-\beG A_0^2)\dot{H}
+8\beG^2\Mpl^2HA_0\dot{A}_0
\right.\notag\\
&&\left.
+32\beG^4H^2A_0^4
-\bem\beG^2MA_0^3\tp-2\beG^2(4\Mpl^2H^2+\tp^2)A_0^2\right]\,.
\label{csmodel}
\ea
Let us derive an approximate expression of $c_{s2}^2$ 
for the coupling $\beta_m^2$ smaller than the order 1.
Eliminating the terms $\dot{H}$, $\dot{A}_0$, $A_0$ 
in Eq.~(\ref{csmodel}) by using the background 
equations of motion, we find 
\be
c_{s2}^2 =1-\frac{2\beta_G \dot{\phi}^2}
{(\beta_A M^2+6\beta_G H^2)M_{\rm pl}^2}
+{\cal O}(\beta_m^2 )\,.
\label{cses}
\ee
Since $\dot{\phi}^2$ is smaller than the orders 
$H^2 M_{\rm pl}^2$ and $M^2 M_{\rm pl}^2$ 
during inflation, 
the condition $c_{s2}^2 > 0$ can be satisfied 
for $|\beta_G| \lesssim |\beta_A|$.

In Fig.~\ref{fig2}, we show the evolution of $q_t, q_v, q_s$ 
and $c_{t}^2, c_{v}^2, c_{s2}^2$ for the same model parameters 
and initial conditions as those used in the left panel 
of Fig.~\ref{fig1}. 
As estimated from Eq.~(\ref{qct}), the quantity $q_t$ is 
close to $M_{\rm pl}^2$  during inflation and reheating 
with a small deviation induced by the time variation of $\phi$. 
In the numerical simulation of Fig.~\ref{fig2}, 
the quantity $q_s$ also remains positive with $q_v=1$.
Hence there are no ghosts of tensor, vector, and 
scalar perturbations. 

As we observe in the right panel of Fig.~\ref{fig2}, 
the deviations of $c_t^2$ and $c_v^2$ from 1 are smaller 
than the order 0.1, so there are no Laplacian instabilities of 
tensor and vector perturbations. 
The scalar propagation speed squared 
$c_{s2}^2$ deviates from 1 in the transient regime from inflation to reheating. 
This comes from the fact that $\dot{\phi}^2$ reaches a 
maximum around the end of inflation.
In the numerical simulation of Fig.~\ref{fig2}, the peak value 
of $|\dot{\phi}|$ is about $0.435MM_{\rm pl}$ with 
$\phi\simeq0.255\Mpl$ and 
the Hubble expansion rate 
$H \simeq 0.187M$ around $Mt \simeq 116$, 
in which case the analytic estimation (\ref{cses}) gives 
$c_{s2}^2 \simeq 0.88$. This exhibits good agreement 
with the minimum value of $c_{s2}^2$ seen in 
Fig.~\ref{fig2}. After the onset of reheating, the term 
$6\beta_G H^2$ becomes negligible relative 
to $\beta_A M^2$, and hence $c_{s2}^2$ approaches 
to 1 according to the relation 
$c_{s2}^2 \simeq 1-2\beta_G \dot{\phi}^2
/(\beta_A M^2 M_{\rm pl}^2)$ with the damped 
oscillation of $\dot{\phi}$. 
Thus, the scalar perturbations are free from Laplacian 
instabilities for the model parameters used in 
Fig.~\ref{fig2}.

\section{Conclusions}
\label{concludesec}

In this paper, we have studied cosmological implications 
of SVT theories with nonlinear derivative scalar 
and vector-field interactions and nonminimal couplings to gravity \cite{Heisenberg18}. 
In Sec.~\ref{modelsec}, we obtained the background equations of 
motion on the flat FLRW spacetime for $U(1)$ broken SVT theories given by 
the action (\ref{action}). The time-dependent scalar field $\phi(t)$ 
as well as the temporal vector component $A_0(t)$ contribute 
to the background cosmological dynamics relevant to the physics of 
cosmic acceleration. The $U(1)$ broken SVT theories contain 
six propagating DOFs-- two scalar modes, two transverse 
vector modes, and two tensor polarizations. 

In Sec.~\ref{tensorsec}, we derived conditions for the absence 
of ghosts and Laplacian instabilities in the tensor sector.
The two polarized states of tensor perturbations have the propagation speed $c_t$ given by Eq.~(\ref{ct}).
Applying these results to the late-time cosmology, we showed that 
the functions $f_4$ and $f_5$ are restricted to be of the form (\ref{f45con})
for the realization of $c_t$ equivalent to 
that of light. We also computed the leading-order primordial 
power spectrum of tensor perturbations generated during inflation, 
see Eq.~(\ref{Ptf}).

In Sec.~\ref{vectorsec}, we considered the vector perturbations $Z_i$ 
and $V_i$ arising from the spatial part of $A_{\mu}$ and the shift 
$V_i$ in the metric, respectively, and obtained their second-order 
actions of the form (\ref{Sv}). 
On using the equation of motion for the non-dynamical field $V_i$, 
the final action (\ref{Sv2}) of vector perturbations contains two 
propagating fields $Z_1$ and $Z_2$ with the same propagation 
speed squared $c_v^2$ given by Eq.~(\ref{cv}).
In the small-scale limit where the vector-field mass squared is 
irrelevant to the dynamics of perturbations, there are neither ghost 
nor Laplacian instabilities under the conditions (\ref{qvcvcon}). 

In Sec.~\ref{scalarsec}, we derived the quadratic action of scalar perturbations by considering 
metric perturbations $\alpha, \chi$
in the flat gauge, scalar perturbations $\delta A, \psi$ arising 
from the temporal and spatial components of $A_{\mu}$ respectively, 
and the scalar-field  perturbation $\delta \phi$. 
After integrating out the non-dynamical fields $\alpha, \chi, \delta A$, 
the action of dynamical perturbations $\psi$ and $\delta \phi$ is 
expressed in the form (\ref{Ss2}).
In the small-scale limit, the no-ghost conditions of scalar 
perturbations are given by Eqs.~(\ref{ng1}) and (\ref{ng2}).
We also derived the two different propagation speed squares 
$c_{s1}^2$ and $c_{s2}^2$, both of which need to be positive 
to avoid small-scale Laplacian instabilities. 
We found that intrinsic vector modes do not affect no-ghost conditions of scalar perturbations, 
but they can modify the values of 
$c_{s1}^2$ and $c_{s2}^2$.

In Sec.~\ref{apsec}, we constructed a concrete model of inflation 
in the framework of SVT theories characterized by the functions 
(\ref{model}). The nonvanishing coupling $\beta_m$ gives rise to 
a kinetic mixing between $\dot{\phi}$ and $A_0$, 
so that the amount of inflation is modified due to the change of 
the slow-roll parameter $\epsilon=-\dot{H}/H^2$.
This property was numerically confirmed for the 
inflaton potential of the $\alpha$-attractor model, 
see Fig.~\ref{fig1}.
We also showed that, under certain bounds of the 
coupling constants $\beta_m, \beta_G, \beta_A$, 
this model can satisfy all the no-ghost and 
stability conditions of tensor, vector, and scalar perturbations 
during inflation and reheating.

It will be of interest to compute inflationary observables relevant to CMB 
temperature anisotropies for the model proposed in this paper.
In particular, the contribution of vector perturbations to CMB 
temperature anisotropies is one of distinguished features of 
SVT theories. The coupling between the scalar and vector fields can also
modify observational predictions of standard slow-roll inflation, 
e.g., the scalar spectral index and the tensor-to-scalar ratio.
Moreover, it will be interesting to apply SVT theories to dark energy 
and estimate observables associated with the background and 
perturbations. 
These issues are left for future works.

\section*{Acknowledgements}

LH thanks financial support from Dr.~Max R\"ossler, 
the Walter Haefner Foundation and the ETH Zurich
Foundation.  
RK is supported by the Grant-in-Aid for Young 
Scientists B of the JSPS No.\,17K14297. 
ST is supported by the Grant-in-Aid 
for Scientific Research Fund of the JSPS No.~16K05359 and 
MEXT KAKENHI Grant-in-Aid for 
Scientific Research on Innovative Areas ``Cosmic Acceleration'' (No.\,15H05890). 
ST thanks warm hospitality to ETH-ITS Zurich 
in which a part of this work was done. 

\appendix

%
\section{Coefficients in the second-order action of scalar perturbations}

The coefficients $D_{1,\cdots,10}$ and $w_{1,\cdots,8}$ 
appearing in Eqs.~(\ref{Lphi})-(\ref{LGP}) are given by
\ba
&&
D_1=\frac{1}{2}\left(f_{2,X_1}+\tp^2 f_{2,X_1X_1}+\tp A_0 f_{2,X_1X_2}
+\frac{A_0^2}{4}f_{2,X_2X_2}\right)\,,
\notag\\
&&
D_2=-\frac12 f_{2,X_1}\,,
\notag\\
&&
D_3=\left( 3f_{4,\phi\phi}+3HA_0f_{5,\phi\phi} \right)\dot{H}
-\frac12\left(f_{2,X_1\phi}+\tp^2f_{2,X_1X_1\phi}+\tp A_0f_{2,X_1X_2\phi}
+\frac{A_0^2}{4}f_{2,X_2X_2\phi}\right)\ddot{\phi}
\notag\\
&&\hspace{0.9cm}
+\frac{H^3A_0(9f_{5,\phi\phi}+A_0^2f_{5,X_3\phi\phi})}{2}
+3H^2\left[2f_{4,\phi\phi}+A_0^2f_{4,X_3\phi\phi}
+\frac{\dot{A_0}(f_{5,\phi\phi}+A_0^2f_{5,X_3\phi\phi})}{2}\right]
\notag\\
&&\hspace{0.9cm}
-3H\left[\frac{\tp}{2}f_{2,X_1\phi}+\frac{A_0(f_{2,X_2\phi}+4f_{3,\phi\phi})}{4}
-A_0\dot{A_0}f_{4,X_3\phi\phi}\right]
-\frac{\tp^2(2f_{2,X_1\phi\phi}+\dot{A_0}f_{2,X_1X_2\phi})}{4}
\notag\\
&&\hspace{0.9cm}
-\frac{\tp A_0}{4}\left[f_{2,X_2\phi\phi}
+\frac{\dot{A_0}(4f_{2,X_1X_3\phi}+f_{2,X_2X_2\phi})}{2}\right]
-\dot{A_0}\left(f_{3,\phi\phi}-A_0^2\tilde{f}_{3,\phi\phi}
+\frac{f_{2,X_2\phi}+A_0^2f_{2,X_2X_3\phi}}{4}\right)+\frac{f_{2,\phi\phi}}{2}\,,
\notag\\
&&
D_4=-\tp^3 f_{2,X_1X_1}-\frac{\tp^2A_0f_{2,X_1X_2}}{2}-
\tp \left( f_{2,X_1}-A_0^2f_{2,X_1X_3} \right) 
+3 H^2 A_0 \left( f_{5,\phi}-A_0^2f_{5,X_3\phi} \right)
+6H \left(f_{4,\phi}-A_0^2f_{4,X_3\phi} \right)
\notag\\
&&\hspace{0.9cm}
+\frac{A_0(f_{2,X_2}+A_0^2f_{2,X_2X_3}+4f_{3,\phi}-4A_0^2\tilde{f}_{3,\phi})}{2}\,,
\notag\\
&&
D_5=H^3A_0^3 \left(f_{5,X_3\phi}+A_0^2f_{5,X_3X_3\phi} \right)
+3H^2\left[\tp A_0 \left( f_{5,\phi\phi}-A_0^2f_{5,X_3\phi\phi} \right)
+2 \left(f_{4,\phi}+A_0^4f_{4,X_3X_3\phi} \right)\right]
\notag\\
&&\hspace{0.9cm}
+6H\left[\tp \left(f_{4,\phi\phi}-A_0^2f_{4,X_3\phi\phi} \right)
-A_0^3 \left( \tilde{f}_{3,\phi}+f_{3,X_3\phi} \right) \right] 
+2\tp A_0 \left( f_{3,\phi\phi}-A_0^2\tilde{f}_{3,\phi\phi} \right)
-\tp^2f_{2,X_1\phi}+A_0^2f_{2,X_3\phi}+f_{2,\phi}\,,
\notag\\
&&
D_6=-2\left(f_{4,\phi}+HA_0f_{5,\phi}\right)\,,
\notag\\
&&
D_7=-H^2A_0 \left( 3f_{5,\phi}+A_0^2f_{5,X_3\phi} \right)
-2H \left( f_{4,\phi}+2A_0^2f_{4,X_3\phi}-\tp A_0f_{5,\phi\phi} \right)
+\tp \left(f_{2,X_1}+2f_{4,\phi\phi} \right)+\frac{A_0(f_{2,X_2}+4f_{3,\phi})}{2}\,,
\notag\\
&&
D_8=-\frac{2\tp D_1+D_4+3HD_6}{A_0}\,,
\notag\\
&&
D_9=-\frac{D_5}{A_0}-2H^3A_0^2f_{5,X_3\phi}
+6H^2\left(\tp f_{5,\phi\phi}+\frac{f_{4,\phi}-A_0^2f_{4,X_3\phi}}{A_0}\right)
+\frac{6H\tp f_{4,\phi\phi}}{A_0}
+\frac{2f_{2,\phi}-2\tp^2 f_{2,X_1\phi}-\tp A_0 f_{2,X_2\phi}}{2A_0}\,,
\notag\\
&&
D_{10}=-2\dot{H}f_{5,\phi}-H^2 \left(3f_{5,\phi}+A_0^2f_{5,X_3\phi} \right)
-2HA_0 \left(2f_{4,X_3\phi}+\dot{A_0}f_{5,X_3\phi} \right)
-2\dot{A_0}f_{4,X_3\phi}+2f_{3,\phi}+\frac{f_{2,X_2}}{2}\,,
\ea
and 
\ba
&&
w_1=-H^2A_0^3 \left( f_{5,X_3}+A_0^2f_{5,X_3X_3} \right)
-2H\left[\tp A_0 \left( f_{5,\phi}-A_0^2f_{5,X_3\phi} \right)
+2f_4+2A_0^4f_{4,X_3X_3}\right]\,,
\notag\\
&&\hspace{0.9cm}
-2\tp \left( f_{4,\phi}-A_0^2f_{4,X_3\phi} \right)
+2A_0^3 \left( \tilde{f}_3+f_{3,X_3} \right)\,,
\notag\\
&&
w_2=w_1+2Hq_t-\tp D_6\,,
\notag\\
&&
w_3=-2A_0^2q_v\,,
\notag\\
&&
w_4=-\frac{H^3A_0^3(9f_{5,X_3}-A_0^4f_{5,X_3X_3X_3})}{2}
-3H^2\left(2f_4+2A_0^2f_{4,X_3}+A_0^4f_{4,X_3X_3}-A_0^6f_{4,X_3X_3X_3}\right)
\notag\\
&&\hspace{0.9cm}
+\frac{3H^2\tp A_0(f_{5,\phi}+4A_0^2f_{5,X_3\phi}-A_0^4f_{5,X_3X_3\phi})}{2}
-3H\tp \left(2f_{4,\phi}-4A_0^2f_{4,X_3\phi}+A_0^4f_{4,X_3X_3\phi} \right)
\notag\\
&&\hspace{0.9cm}
+3HA_0^3\left[\tilde{f}_3+f_{3,X_3}-A_0^2 \left( \tilde{f}_{3,X_3}+f_{3,X_3X_3} \right)\right]
-\tp A_0 \left[3f_{3,\phi}-A_0^2\left( \tilde{f}_{3,\phi}+f_{3,X_3\phi} \right)
+A_0^4\tilde{f}_{3,X_3\phi}\right]
\notag\\
&&\hspace{0.9cm}
+\frac12\left[\tp^4 f_{2,X_1X_1}+\tp^2 \left( f_{2,X_1}-2A_0^2f_{2,X_1X_3} \right)
+A_0^4f_{2,X_3X_3}-\frac{3\tp A_0 f_{2,X_2}}{2}\right]\,,
\notag\\
&&
w_5=w_4-\frac{3H(w_1+w_2)}{2}-3H^2\tp A_0^3f_{5,X_3\phi}
+3H\tp \left(f_{4,\phi}-2A_0^2f_{4,X_3\phi} \right)
+2\tp A_0 \left(f_{3,\phi}-A_0^2\tilde{f}_{3,\phi} \right)
\notag\\
&&\hspace{0.9cm}
+\frac{\tp}{2}\left[\tp A_0^2\left(2f_{2,X_1X_3}+\frac14f_{2,X_2X_2}\right)
-\tp^3f_{2,X_1X_1}+A_0^3f_{2,X_2X_3}-\tp f_{2,X_1}+A_0f_{2,X_2}\right]\,,
\notag\\
&&
w_6=-\frac{2w_1-w_2-2\tp D_6+8Hf_4}{A_0}\,,
\notag\\
&&
w_7=-2\dot{H} \left(2f_{4,X_3}+HA_0f_{5,X_3} \right)
-H^2\left[\frac{\tp(3f_{5,\phi}+A_0^2f_{5,X_3\phi})}{A_0}
+\dot{A_0} \left(f_{5,X_3}+A_0^2f_{5,X_3X_3} \right)\right]\notag\\
&&\hspace{0.9cm}
-4H \left( \tp f_{4,X_3\phi}+A_0\dot{A_0}f_{4,X_3X_3} \right)
+2\dot{A_0} \left(\tilde{f}_3+f_{3,X_3} \right)+\frac{\tp(4f_{3,\phi}+f_{2,X_2})}{2A_0}\,,
\notag\\
&&
w_8=3Hw_1-2w_4-\tp D_4\,.
\ea
%


\end{document}